\documentclass[pra]{revtex4}
\usepackage{amsmath}
\usepackage{graphicx}
\usepackage{color}
\newcommand{\be}[1]{\begin{equation}\label{#1}}
\newcommand{\ee}{\end{equation}}
\newcommand{\bea}[1]{\begin{eqnarray}\label{#1}}
\newcommand{\eea}{\end{eqnarray}}
\newcommand{\no}{\nonumber \\}
\newcommand{\Fig}[1]{Fig.(\ref{#1})}
\newcommand{\Eq}[1]{Eq.(\ref{#1})}

\def\a0{{\alpha_0}}

\def\da0{{\dot{\alpha}_0}}

\newcommand{\adag}{{a^\dagger}}
\newcommand{\ket}[1]{|#1\rangle}
\newcommand{\bra}[1]{\langle #1|}

\def\myover#1{\myoverDefn#1}
\def\myoverDefn#1#2{\hbox{\space \raise-2mm\hbox{$\textstyle{#1} \atop \scriptstyle{#2}$} }}

\def\ibar{{\bar{i}}}

\def\np{{n_p}}
\def\np0{{n_{p0}}}

\def\apdag{a_p^\dagger}
\def\asdag{a_s^\dagger}
\def\aibardag{a_\ibar^\dagger}
\def\ap{a_p}
\def\as{a_s}
\def\aibar{a_\ibar}

\begin{document}
\title{Spontaneous parametric down conversion with a depleted pump\\as an analogue for black hole evaporation/particle production}
\author{P.M. Alsing and M.L. Fanto}
\affiliation{Air Force Research Laboratory, Rome, NY, 13441}
\date{\today}

\begin{abstract}
We present an analytical formulation of the recent one-shot decoupling model of Br\`adler and Adami [arXiv:1505.0284] and compute the resulting ``Page Information" curves, for the reduced density matrices for the evaporating black hole internal degrees of freedom, and emitted Hawking radiation pairs entangled across the horizon. We argue that black hole evaporation/particle production has a very close analogy to the laboratory process of spontaneous parametric down conversion, when the pump is allowed to deplete.
\end{abstract}
\maketitle

\section{Introduction}\label{Intro}
Recently there has been renewed interest in the use of a \textit{trilinear Hamiltonian} to phenomenologically and  explicitly unitarily model the evaporation of quantized internal degrees of freedom (treated as scalar bosons) of a black hole (BH) while commensurately generating Hawking radiation pairs \cite{Nation_Blencowe:2010,Alsing:2015,Bradler_Adami:2015}, one of which falls inward behind the BH horizon and the other, observed mode that has a thermal distribution, at least for early evolution times. The goal of representing the internal degrees of freedom of the BH as quantized is to be able to study both the evaporation of the BH as well as the effect on the thermality of the emerging Hawking particles. In the work of Nation and Blencowe \cite{Nation_Blencowe:2010} and Alsing \cite{Alsing:2015} a single Hawking radiation pair mode was considered, that coupled to an internal quantized mode of the BH. Both models reproduced  information curves conjectured by Page \cite{Page:1993a,Page:1993b} which predicted that the difference between the von Neumann entropy of an effective thermal density matrix for the outgoing Hawking radiation and the actual entropy, the so-called \textit{Page information} $I = S(\rho_{thermal}) - S(\rho)$, would exhibit the following behavior, namely (i) for early evolution times $I$ would be flat and nearly zero, since the BH is essentially un-depleted, and the emitted Hawking radiation has a thermal nature, and (ii) for long times, when the number of emitted Hawking particles is on the order of the remaining number of particle in the BH, $I$ would begin to rise rapidly as the BH evaporates, and the emitted Hawking radiation deviates from a thermal state.

In a recent paper, Br\`adler and Adami \cite{Bradler_Adami:2015} have expanded upon the work Nation and Blencowe \cite{Nation_Blencowe:2010} and Alsing \cite{Alsing:2015} and developed a one-shot decoupling model for BH evaporation, in which the quantized modes of the BH are emitted sequentially in time into the out-going and infalling radiation modes \footnote{A similar, related open-systems model was advocated for in the closing discussion section of Alsing \cite{Alsing:2015}}. The authors use the method of lattice paths to compute the von Neumann entropy of the reduced density matrix for the BH and show that its resulting time evolution is again qualitatively similar to that predicted by Page, in which information emerges from the BH at a time when the BH has evaporated to roughly half its initial size (population). The new feature here is the use of a sequence of temporally emitted Hawking radiation modes (vs one) to more realistically model the creation of a  train of Hawking pairs, each of which interacts with the BH quantized internal degree of freedom for some finite amount of time before the next subsequent Hawking pair emission event. Such a model conforms to the suspected physical generation process of Hawking radiation as suggested by many past \cite{Boulware:1976,Gerlach:1977} and recent works \cite{Mathur:2009}. As discussed by Mathur \cite{Mathur:2009} in his informative, pedagogical introduction to the BH Information Problem, Hawking pairs are created in a region of curvature distortion near the BH horizon, and subsequently propagate away from the region of generation, while the BH horizon shrinks in radius. As such, the next generated pair does not interact with the previously generated pairs, and is coherent only with the BH internal degrees of freedom for a time on the order of average time between emission events. This is what the one-shot decoupling model of Br\'adler and Adami \cite{Bradler_Adami:2015} seeks to capture.

The goal of this present paper is to analytically expand upon the one-shot decoupling model based on the trilinear Hamiltonian of Alsing \cite{Alsing:2015}, and discussed in detail in the next section. The new aspect of this present work is the development of analytic approximations for the probabilities to the quantized BH / Hawking radiation state which are subsequently used to generate Page information curves, without the need for excessive computational resources. We further argue that the presented model of combined BH evaporation/Hawking pair production has very close analogies to the laboratory process of spontaneous parametric down conversion (SPDC) \cite{Gerry:2004,Agarwal:2013} with a finite (vs infinite) BH `pump' source capable of depletion (the focus of \cite{Alsing:2015}).

The outline of this paper is a follows.
In Section \ref{HawkRad} we describe the trilinear Hamiltonian used to the manifestly, unitarily model the BH evaporation/Hawking pair production, and
in Section \ref{OneShot} review the one-shot decoupling model of Br\'adler and Adami  \cite{Bradler_Adami:2015}.
In Section \ref{NumericalResults} we develop analytic approximations to the probability distributions for the reduced density matrix of the state of the BH, and for the state of the emitted Hawking radiation. Based on these approximations to the probabilities, we develop Page Information curves which illustrate the predictions conjectured by Page \cite{Page:1993b}.
In the final discussion section
Section \ref{Discussion} we make the analogy of the this one-shot decoupling to the process of SPDC
generated by a laser powered by a battery of finite capacity.

\section{Review of trilinear Hamiltonian model}\label{HawkRad}
In the work by Alsing \cite{Alsing:2015} the evaporating BH hole was modeled by the manifestly unitary process given by the following trilinear Hamiltonian
\be{H:eqn}
H_{p,s,\ibar} = r \left(
                         a_p a^\dagger_s a^\dagger_\ibar
                       + a^\dagger_p a_s a_\ibar
                  \right),
\ee
where $\apdag, \, \ap$ are the creation and annihilation operators for the BH quantized, internal degrees of freedom, and $\asdag, \, \as$ and $\aibardag, \, \aibar$ are the creation and annihilation operators for the outgoing (observed) and in-falling (unobserved, behind the BH horizon) modes of the emitted Hawking radiation.
$H_{p,s,\ibar}$ also models the process of parametric down conversion with a depleted pump source \cite{Walls:1970,Bonifacio:1970}.
Following the notation of \cite{Alsing:2015} we use the subscript notation $p$ to denote the BH `pump' soruce, and the labels $s$ and $\ibar$ to denote emitted the `signal' and `idler' modes respectively. Without loss of generality we take the signal mode $s$ to be the particle and the idler mode $\ibar$ as the anti-particle emitted externally and internally, respectively to the BH horizon.
Defining the initial state of the combined system as
\be{psi0}
\ket{\psi(0)} = \ket{\np0}_p\ket{0}_s\ket{0}_\ibar \equiv  \ket{\np0}_p\ket{0}_{s,\ibar},
\ee
where $\np0\gg 1$ is the initial boson occupation number for the BH.
By solving the Schrodinger equation for the state
$\ket{\psi(\tau)} = \sum_{n=0}^{\np0} \, c_n(\tau)\, \ket{\np0-n}_p\ket{n}_s\ket{n}_\ibar$
$\equiv\sum_{n=0}^{\np0} \, c_n(\tau)\, \ket{\np0-n}_p\ket{n}_{s,\ibar}$
in both the short-time ($\np0\gg n$) and long-time ($\np0,n\gg 1$) one obtains the
solutions \cite{Alsing:2015}
\bea{psi:soln}
c_n^<(z,\tau) &=& \left[ (1-z) \, z^n \right]^{1/2},
\hspace{1.075in} 0 \le z \le z^*,
\quad z^* \approx 0.506407 \hspace{1em} \textrm{for} \hspace{1em} \np0 \gg 1,\\
c_n^>(z,\tau) &=& \left[ c_n^<(f(z)/(1+f(z)),\tau)\right]^{1/2},
\hspace{.25in} z^* \le z \le 1, \\
f(z) &=& 4 e^{-\pi}\,\left(
                            \frac{1+\sqrt{z}}{1-\sqrt{z}}
                      \right)
                            = 4 e^{-\pi} e^{2\tau},\qquad
                            z = \tanh^2(\tau), \quad \tau = \sqrt{\np0}\,r\,t \label{z:defn}.
\eea
In the shorttime limit ($\np0 \gg n$) the BH `pump' source is essentially undepleted, and one can factor out the state $\ket{\np0}_p$ using the approximation $\np0-n\approx\np0$, to yield the well studied two-mode squeezed state $\ket{\phi}_{s,\ibar}$
form of the emitted Hawking radiation
\bea{psi:shorttime}
\ket{\psi(z,\tau)}^< &\approx& \ket{\np0}_p\,\ket{\phi}_{s,\ibar}, \qquad z < z^*,\\
\ket{\phi}_{s,\ibar}^{(sqzd)} &=& \sqrt{\frac{1-z}{1-z^{\np0+1}}}\,
                         \sum_{n=0}^{\np0} z^{n/2} \, \ket{n}_{s,\ibar} \approx
 \sqrt{1-z}\,\sum_{n=0}^{\infty}\, z^{n/2} \ket{n}_{s,\ibar}, \; (\np0\rightarrow\infty),
\eea
where $\ket{n}_{s,\ibar} = (\asdag,\aibardag)^n/n!\,\ket{0}_{s,\ibar}$.

In \cite{Alsing:2015} the initial number of bosons $\np0$ in the BH was taken to be finite, though sufficiently large
$\np0\gg 1$ so that without loss of generality all upper limits of summations could safely be taken as infinity
(though in numerical computations, a finite value of $\np0$ was utilized).
In this work we will be especially careful to  keep track of the upper limits of all summations, with $\np0$ large
but finite so that \Eq{psi:soln} is formally given by
\be{cn:shorttime}
c_n^<(z,\tau) = \left[ \frac{(1-z)}{(1-z^{\np0+1})} \, z^n \right]^{1/2} \equiv \sqrt{p_n^<(z,\tau)},
\ee
such that
\be{pn:shorttime}
\sum_{n=0}^{\np0} p_n^<(z,\tau) =1.
\ee

A physical motivation for utilizing the trilinear Hamiltonian \Eq{H:eqn} to model BH evaporation/Hawking pair creation is as follows. The standard approach to modeling Hawking pair creation taken by Hawking and numerous subsequent authors is to treat the gravitation field of an eternal BH (typically taken as Schwarzschild, without loss of generality) as a classical background field, to which a quantized, scalar boson field is coupled \cite{Hawking:1975}. The subsequent state of emitted Hawking radiation pairs was shown by Unruh \cite{Unruh:1976} to be a two mode squeezed state $\ket{\phi}^{(sqzd)}_{s,\ibar}$  \cite{Gerry:2004,Agarwal:2013}, which upon tracing out of the (idler) mode $\ibar$ which falls behind the horizon, becomes a thermal state \cite{Yurke_Potasek:1987}.
Very early on, researchers such as Boulware \cite{Boulware:1976} and Gerlach \cite{Gerlach:1977}, and more recently by authors such as Stojkovic \textit{et. al.} \cite{Stojkovic:2006:2015} and Alberghi \textit{et. al.} \cite{Alberghi:2001}, considered BH evaporation from a collapsing thin shell of matter. Here, the BH matter is still treated classically, and is characterized by a time dependent, shrinking horizon radius. The Hamiltonian derived for a scalar boson field coupled to this classical gravitation field (see \cite{Stojkovic:2006:2015,Alberghi:2001}) leads to a quantized harmonic oscillator with a (exponentially) time varying frequency. Such a Hamiltonian can be written in the form \cite{Pedrosa:1987:1989,Alsing_Rice:2015}
$H = 1/2\,r\,\xi(\tau)\,(\adag^2 + a^2)$ involving the generators of $su(1,1)$ and well known to generate single mode squeezed states \cite{Gerry:2004,Agarwal:2013}.
Here $\xi(\tau)$ represents the classical (i.e. c-number) `driving field' of the collapsing shell of matter.
The single mode squeezed state arises from the coupling of a single quantized scalar boson field to the classical gravitational field. If two bosons were coupled to the field, or a single, complex boson field, the Hamiltonian would be of the form
$H = 1/2\, r\, \xi(\tau)\, (\adag_s\,\adag_{\ibar} + a_s\,a_{\ibar})$ where we have labeled the correlated emitted Hawking pair as a signal and idler mode. Again, the gravitation field shows up as a classical c-number driving field.
In quantum optics, such semi-classical models are familiar, where the occupation number of a strong driving pump laser is so large, that for all intent and purposes concerning the subsequent particle statistics, it is eminently reasonable to consider it as a classical c-number field \cite{Gerry:2004,Agarwal:2013}. The next logical step, in order to incorporate the quantum statistics of the pump, is to replace the classical driving field $\xi(\tau)$ by quantized mode $a_p$ and $\adag_p$.
This is the trilinear Hamiltonian \Eq{H:eqn} used in this work
(and in \cite{Nation_Blencowe:2010,Alsing:2015,Bradler_Adami:2015}).
Since at present, there does not exist a well accepted description of a quantized gravitational field (Schwarzschild, collapsing shell, etc\ldots), this trilinear Hamiltonian, though physically motivated and eminently reasonable, must be taken to be at best phenomenological. It's main advantage is that it manifestly unitary while capturing the essential qualitative features of generating across-the-horizon entangled Hawking radiation pairs, as well as serving as a simple model for BH evaporation.

\section{One-shot decoupling model}\label{OneShot}
The work of Br\'adler and Adami \cite{Bradler_Adami:2015} generalizes the process used in
Section \ref{HawkRad} to the more physically relevant one-shot decoupling state in which over some period $\Delta{\tau}$ a Hawking signal/idler pair is generated by the curvature in the vicinity of the instantaneous BH horizon (with radius formally proportional to $n_p(\tau)$, the instantaneous BH occupation number), which then travels away from the region of generation, never to interact with the BH again,
as discussed by \cite{Gerlach:1977,Mathur:2009}
(and as suggested in the summary/conclusion of \cite{Alsing:2015}).
Such a model is physically analogous to spontaneous parametric down conversion (SPDC) in a nonlinear crystal of finite length (as is typical in laboratory experiments) in which the entangled (here)  photon pairs are generated inside the crystal (as some random spatial position). Upon exiting the crystal, the signal/idler pair no longer participate in the SPDC process described by the Hamiltonian \Eq{H:eqn}.
Thus the generated signal/idler modes emerge as a temporal sequence of emitted entangled pairs.
Br\'adler and Adami model this using the initial state
\be{Psi:BA}
\ket{\Psi(0)} = \ket{\np0}_p\prod_{k'=1}^N \ket{0}_{s_{k'},\ibar_{k'}}=
\ket{\np0}_p\otimes
\ket{0}_{s_1,\ibar_1}\otimes
\ket{0}_{s_2,\ibar_2}\otimes\ldots\otimes
\ket{0}_{s_N,\ibar_N},
\ee
where $\tau = N \Delta\tau$, and $N$ is the number of time slices.
The evolution of the state $\ket{\psi(0)}$ in \Eq{Psi:BA} is given by \cite{Bradler_Adami:2015}
\be{U:T}
\ket{\psi(\tau)} = U(\tau,0) \ket{\Psi(0)}
= \mathcal{T} e^{-i\int d\tau' \, H_{p,s,\ibar}(\tau')}\,\ket{\Psi(0)}
\approx \prod_{k=1}^N \,
e^{-i H_{p,s_k,\ibar_k}\Delta\tau}\,\ket{\np0}_p\,
\prod_{k'=1}^N\,\ket{0}_{s_{k'},\ibar_{k'}},
\ee
where ${\mathcal{T}}$ is the time-ordered product and in the second equality we have used a simplified version of the Trotter expansion valid for $N$ small time
slices of size $\Delta\tau$, with $U_{p,k} = e^{-i H_{p,s_k,\ibar_k}\Delta\tau}$ acting on modes $p$ and $(s_k,\ibar_k)$.

After the first time slice, the wave function is
\bea{Psi:1}
\ket{\Psi(1)} &=&U_{p,1}\,\ket{\Psi(0)} =  \sum_{n_1=0}^\np0 \sqrt{p_{n_1}^{(n)}(z)}\,\,
     \ket{\np0-n_1}_p\ket{n_1}_{s_1,\ibar_1}\otimes
     \prod_{k'=2}^N\,\ket{0}_{s_{k'},\ibar_{k'}}, \qquad z \ll z^*, \\\label{Psi:1:exact}
   &\equiv& \sum_{n_1=0}^\np0 \sqrt{p_{n_1}^{(\np0)}(z)}\,\,
     \ket{\np0-n_1}_p\ket{n_1}_{1},
     \quad p_{n_1}^{(\np0)}(z) = \frac{(1-z)}{(1-z^{\np0+1})}\,z^{n_1},
     \quad \sum_{n_1=0}^\np0\,p_{n_1}^{(\np0)}=1,
     \\
&\approx&
\ket{\np0}_p\otimes\sum_{n_1=0}^\np0 \sqrt{p_{n_1}^{(\np0)}(z)} \,\, \ket{n_1}_{1},     \qquad \np0\gg n_1, \\
&\equiv&
\ket{\np0}_p\otimes \ket{\phi^{(sqzd)}}_{1},\label{Psi:1:approx}
\eea
where $\ket{\phi^{(sqzd)}}_{1}=(1-z)\sum_{n_1=0}^{\np0\rightarrow\infty} z^{n_1} \ket{n_1}_1$ is two-mode
signal/idler emittted Hawking radiation state.
The emitted Hawking signal/idler pairs are approximately squeezed for early time $z<z^*$, since for long time evolution the exact state in \Eq{Psi:1:exact} does not factorize as in case of the
short time state \Eq{Psi:1:approx}.
Note that notation $p_{n_1}^{(n)}$ indicates the probability that $n_1$ particles are emitted into the Hawking radiation signal/idler mode when there were initially $\np0$ particles in the BH `pump' mode.
Henceforth, we shall
denote  $\ket{n_i}_{i}\equiv \ket{n_i}_{s_i,\ibar_i}$, drop the argument $z$ on the probabilities,
and leave implied the unoccupied vacuum signal/idler states $\ket{0}_{s_{k'},\ibar_{k'}}$ for $k'$ greater than the current timeslice considered. From \Eq{cn:shorttime} and \Eq{pn:shorttime} the state $\ket{\Psi(1)}$ is clearly normalized to unity.

To illustrate the notation we will employ, it is instructive to write down the
wavefunction at after the second emission event
\bea{Psi:2}
\ket{\Psi(2)} &=& U_{p,2}\,\ket{\Psi(1)}
              = \sum_{n_1=0}^\np0 \, \sum_{n_2=0}^{\np0-n_1}\,
              \sqrt{p_{n_1}^{(\np0)}\,p_{n_2}^{(\np0)-n_1}}\,
                 \ket{(\np0-n_1)-n_2}_p\,\ket{n_1}_{1}
                                         \,\ket{n_2}_{2}, \\ \label{Psi:2:exact}
&\approx&
 \ket{\np0}_p\otimes
\sum_{n_1=0}^\np0 \,  \sqrt{p_{n_1}^{(\np0)}}\,\ket{n_1}_{1}\otimes
\sum_{n_2=0}^{\np0-n_1}\,\sqrt{p_{n_2}^{(\np0-n_1)}}\,\ket{n_2}_{2}, \quad \np0 \gg n_1, n_2,
     \quad p_{n_2}^{(\np0-n_1)}(z) = \frac{(1-z)}{(1-z^{(\np0-n_1)+1})}\,z^{n_2},\no
&\approx&\ket{\np0}_p\otimes\ket{\phi^{(sqzd)}}_1\otimes\ket{\phi^{(sqzd)}}_2,
     \quad \hspace{2.4in} \sum_{n_2=0}^{\np0-n_1}\,p_{n_2}^{(\np0-n_1)}=1. \label{Psi:2:approx}
\eea
The new feature of \Eq{Psi:2} is that the second particle has been emitted into the second signal/idler mode with the only dependence upon mode $1$ being that the initial number of particles in the BH `pump' source is now $\np0-n_1$, where $n_1$ is the number of particles that were emitted into mode $1$ during the first emission event
(note: $n_1\in(1,\ldots,N)\equiv 1:N$).
Again, in the short time limit \Eq{Psi:2:approx} indicates that the emittted Hawking radiation is a succession of independent two-mode squeezed states in modes $1$ and $2$ respectively.

Note that by utilizing a wavefunction $\ket{\Psi(2)}$ we are implicitly assuming a degree of coherency between the pump and the emitted Hawking radiation signal/idler modes, as exhibited in the exact states for $\ket{\Psi(1)}$ and $\ket{\Psi(2)}$ in \Eq{Psi:1} and \Eq{Psi:2} respectively.
The goal of the one-shot decoupling procedure  is to decouple the emitted Hawking radiation modes from the pump at each emission event, while also keeping track of the finite and decreasing nature of the BH quantized degree of freedom $n_p(\tau)$ that arises from the finite, though large, initial occupation number $\np0\gg 1$. In the language of laboratory SPDC, on is making the implicit assumption that the coherency of the BH `pump' source is shorter than the average time between emission events. We return to a discussion of this point in Section \ref{Discussion}.

Since each unitary emission $\{U_{p,i}\}_{i=1:N}$ acts for a short time $\Delta\tau$,
we are continually in the short time regime
$z<z^*$ and each emitted signal/idler Hawking radiation pair is nearly a two-mode squeezed state. However, the occupation number of the BH `pump' mode is continually decreasing, and it is the effect of this finite nature of the `pump' source on the total state that we wish to examine for long times (large $N$) as the BH evaporates.
%
%
Consider the wavefunction $\ket{\Psi(N)}$ after $N$ emitted events given by the generalization of \Eq{Psi:2}
\bea{Psi:N}
\ket{\Psi(N)}
&=&
       \sum_{n_1=0}^\np0 \,
       \sum_{n_2=0}^{\np0-n_1}\,
       \sum_{n_3=0}^{\np0-(n_1+n_2)}\ldots\sum_{n_N=0}^{\np0-(n_1+\ldots+n_{N-1})}\,
        \sqrt{p^{(n)}_{n_1}\,p_{n_2}^{(\np0-n_1)}\,p_{n_3}^{(\np0-n_1-n_2)}\ldots
        p_{n_N}^{(\np0-n_1-\ldots-n_{N-1})}}\,\no
&\times&
 \ket{\np0-(n_1+\ldots+n_N)}_p
 \otimes\prod_{i=1}^N \,\ket{n_i}_{i}, \\
&\approx&
  \ket{\np0}_p\otimes
  \prod_{i=1}^N \,\ket{\phi^{(sqzd)}}_{i}, \qquad \np0 \gg \left.\{n_i\}\right|_{i=1:N}\label{Psi:N:approx}, \\
&\equiv&
  \ket{\np0}_p\otimes\ket{\Phi^{(sqzd)}(N)}.
\eea
By construction we have $\bra{\Psi(N)}\Psi(N)\rangle=1$.

Let us rewrite $\ket{\Psi(N)}$ as follows. We define $j_i = \sum_{m=0}^{i} n_m$ with $j_0\equiv0$. Keeping track of the upper and lower limits on each summation, we obtain the representation
\bea{Psi:j:jN:last}
\ket{\Psi(N)}
&=&
       (1-z)^{N/2}\,\sum_{j_1=0}^\np0 \,
       \sum_{j_2=j_1}^{\np0}\,
       \sum_{j_3=j_2}^{\np0}
       \ldots
       \sum_{j_N=0}^{\np0}\,
       \sqrt{z^{j_N}} \ket{\np0-j_N}_N \otimes
       \prod_{i=1}^N \frac{1}{\sqrt{(1-z^{\np0-j_i}+1)}} \, \ket{j_i-j_{i-1}}_i, \no
&=&
       (1-z)^{N/2}\
       \sum_{j_N=0}^{\np0}  \,\sqrt{z^{j_N}} \, \ket{\np0-j_N}_N\,\otimes
\left[
       \sum_{j_1=0}^{j_N} \,
       \sum_{j_2=j_1}^{j_N}\,
       \sum_{j_3=j_2}^{j_N}
       \ldots
       \sum_{j_{N-1}=j_{N-2}}^{j_N}\,
       \prod_{i=1}^N \frac{1}{\sqrt{(1-z^{\np0-j_{i-1}+1})}} \, \ket{j_i-j_{i-1}}_i,
\right] \qquad \label{Psi:j:jN:first} \\
&\equiv&
(1-z)^{N/2}\,
\sum_{j_{N}=0}^{\np0}\,  \,\sqrt{z^{j_N}} \, \ket{\np0-j_N}_N\otimes\ket{\Phi^{(N)}_{j_N}},   \label{Psi:N}
\eea
where we have defined the unnormalized state $\ket{\Phi^{(N)}_{j_N}}$ by the expression in the large square brackets in
\Eq{Psi:j:jN:first}, and we have pulled the sum over the index $j_N$ to the far left, which alters the limits of the remaining inner nested sums.
$\ket{\Psi^{(N)}_{j_N}}$ describes the emitted Hawking radiation state with exactly $j_N$ particles
(at the $N$th time slice) emitted into $N$ possible distinct signal/idler modes,
which is in general a superposition state over many Fock states whose occupation numbers sum to exactly $j_N$.

We now wish to approximate $\ket{\Phi^{(N)}_{j_N}}$ for large, but finite $\np0\gg 1$.
In \Eq{Psi:j:jN:first} the factors $(1-z^{\np0-j_i+1})^{-1/2}$ are negligibly small for all but $j_i\sim \np0$. Even for $j_i=\np0$ the factor only contributes a $(1-z^{1})^{-1/2} \approx 1+z/2$ for $z\ll 1$, the case we will consider in this work (i.e. weak two-mode squeezed states).
Note the case $j_i = \np0$ corresponds to all $\np0$ particles of the BH emitted in a single burst at time $\tau_i= i\,\Delta\tau$ into mode $(s_i,\ibar_i)$, which constitutes a very low probability event for early times, but not necessarily so for longer times.
Thus, to lowest order, we approximate all the summands by unity for all modes $i=1:N$.
i.e. $(1-z^{\np0-j_i+1})^{-1/2}\approx 1$. Therefore, we have
\bea{Phi:N:approx}
\ket{\Phi^{(N)}_{j_N}} &\approx& \ket{\tilde{\Phi}^{(N)}_{j_N}} =
       \sum_{j_1=0}^{j_N} \,
       \sum_{j_2=j_1}^{j_N}\,
       \sum_{j_3=j_2}^{j_N}
       \ldots
       \sum_{j_{N-1}=j_{N-2}}^{j_N}\,
       \prod_{i=1}^N 
       \, \ket{j_i-j_{i-1}}_i,\no
&=&
\sum_{j_1\le j_2\ldots \le j_{N-2}\le j_{N-1}}^{j_N}\,\,
\prod_{i=1}^N \, \ket{j_i-j_{i-1}}_i,
\eea
with
\be{binomial}
\bra{\tilde{\Phi}^{(N)}_{j_N}} \tilde{\Phi}^{(N)}_{j_N}\rangle =
\left(
           \begin{array}{c}
             j_N+N-1 \\
             j_N
           \end{array}
\right),
\ee
where the binomial factor in \Eq{binomial} counts the number states containing exactly $j_N$ Hawking radiation particles into $N$ signal/idler modes,
i.e. the selection of $j_N+N-1$ objects taken $j_N$ at a time \textit{with} repetitions.
We can also intuitively understand the nested sum in \Eq{Phi:N:approx}
over the dummy indices $j_1\le j_2\ldots \le j_{N-2}\le j_{N-1}$  as the number of lattice points in the `upper diagonal' quadrant (including the diagonal)
of a $N-1$ dimension hypercube with $j_N+1$ lattices points ($0,1,\ldots,j_N$) per dimension.

We normalize the state as
\be{}
\ket{\tilde{\Phi}^{(N)}_{j_N}} \rightarrow \ket{\Phi'^{(N)}_{j_N}} =
\left(
           \begin{array}{c}
             j_N+N-1 \\
             j_N
           \end{array}
\right)^{-1/2}\,
\ket{\tilde{\Phi}^{(N)}_{j_N}},
\qquad
\bra{\Phi'^{(N)}_{\np0-j_N}} \Phi'^{(N)}_{\np0-j_N} \rangle = 1.
\ee
As an example
\bea{Phi:N4:jN2}
\ket{\Phi^{(N=4)}_{j_N=2}} &=&
\left(\ket{2,0,0,0} + \ket{0,2,0,0} +\ket{0,0,2,0} +\ket{0,0,0,2} +\ket{1,1,0,0}
\right. \no
&+&
\left.
 \ket{1,0,1,0} +\ket{1,0,0,1} +\ket{0,1,1,0} +\ket{0,1,0,1} + \ket{0,0,1,1}
\right)/\sqrt{10},
\quad
\left.\left(
           \begin{array}{c}
             j_N+N-1 \\
             j_N
           \end{array}
\right)\right|_{N=4,j_N=2} \hspace{-.5in}= 10,
\eea

Upon replacing $j_N\rightarrow k$ (to simplify notation) as the total number of Hawking radiation particles emitted into $N$ signal/idler modes, we obtain
\bea{Psi:Normalized}
\ket{\Psi(N)}
&\approx&
\sum_{k=0}^{\np0} \sqrt{P_{k}^{(N)}}\, \,\ket{\np0-k}_p \,\ket{\Phi'^{(N)}_{k}},
\quad
\tilde{P}_{k}^{(N)} = (1-z)^N\, z^{k}\,
\left(
       \begin{array}{c}
          k + N-1 \\
          k
        \end{array}
      \right),
\quad
P_{k}^{(N)} = \frac{\tilde{P}_{k}^{(N)}}{\sum_{k'=0}^{\np0}\,\tilde{P}_{k'}^{(N)}}, \qquad
\\
&\approx&
\,\ket{\np0}_p  \otimes \sum_{k=0}^{\np0} \sqrt{P_{k}^{(N)}}\, \,\ket{\Phi'^{(N)}_{k}},
\hspace{0.20in} \left.\np0\gg {j_i}\right|_{i=1:N}, \\
&\myover{{\longrightarrow} {\np0\gg 1}}&
\,\ket{\np0}_p  \otimes \prod_{i=1}^{N} \ket{\phi^{(sqzd)}}_i,
\hspace{0.30in}\qquad \ket{\phi^{(sqzd)}}_i= (1-z) \sum_{n=0}^\infty z^n  \,\ket{n}_i,
\eea
where in the last line, the emitted Hawking radiation signal/idler field modes are in a product of $N$ squeezed states.
\Eq{Psi:Normalized} with probability $P_{k}^{(N)}$  is one of the primary analytic result of this paper, and leads (in the next section) to entropy curves proposed by Page, and discussed in Blencowe and Nation \cite{Nation_Blencowe:2010} and Alsing \cite{Alsing:2015},  and examined numerically for this current one-shot decoupling model in Br\'adler and Adami \cite{Bradler_Adami:2015}.
Note that in the limit $\np0\rightarrow\infty$ we have $\sum_{k=0}^{\infty}\,\tilde{P}_{k}^{(N)}=1$ using the identity
$
\sum_{k=0}^{\infty}\,z^k
\tiny{
\left(
       \begin{array}{c}
          k + N-1 \\
          k
        \end{array}
      \right)
}
= (1-z)^{-N}.
$
It is informative to compare the above state of the emitted Hawking radiation with the separable product of $N$ two-mode squeezed states.

The product of $N$ squeezed states $\ket{\Phi^{(sqzd)}(N)}_{s,\ibar} \equiv \prod_{i=1}^{N} \ket{\phi^{(sqzd)}}_i$ with
$\ket{\phi^{(sqzd)}}_i = (1-z) \sum_{n_i=0}^{\infty} z^{n_i}\ket{n_i}_i$
(where $\ket{n_i}_i \equiv\ket{n_i}_{s_i}\,\ket{n_i}_{{\ibar}_i}$) can be written as
$\ket{\Phi^{(sqzd)}(N)}_{s,\ibar} = \sum_{k=0}^{\infty}\, \sqrt{\tilde{P}_{k}^{(N)}}\, \ket{\Phi_k^{'(N)}}$
where $\tilde{P}_{k}^{(N)} \stackrel{\np0\gg 1}{\rightarrow} P_{k}^{(N)}$.
The density matrix  $\rho^{(sqzd)}_{s,\ibar}$
$=\sum_{k=0}^{\infty}\, \sum_{k'=0}^{\infty}\,\sqrt{\tilde{P}_{k}^{(N)}\,\tilde{P}_{k'}^{(N)}}\,
\ket{\Phi_{k}^{'(N)}}\bra{\Phi_{k'}^{'(N)}}$
of this pure state contains off-diagonal $\sqrt{\tilde{P}_{k}^{(N)}\,\tilde{P}_{k'}^{(N)}}$,  as well as diagonal
matrix elements $\tilde{P}_{k}^{(N)}$, but of course has only a single non-zero eigenvalue of unity, since it is a pure state.
The bipartite state of the BH `pump' mode and all signal/idler modes
$\ket{\Psi(N)} = \sum_{k=0}^{\np0} \sqrt{P_{k}^{(N)}}\, \,\ket{\np0-k}_p \,\ket{\Phi'^{(N)}_{k}}$
in \Eq{Psi:Normalized} has an important difference, even in the limit $\np0\rightarrow\infty$.
In forming the density matrix $\rho_p(N) = Tr_{s,\ibar}[\ket{\Psi(N)}\bra{\Psi(N)}]$ and
$\rho_{s,\ibar}(N) = Tr_{p}[\ket{\Psi(N)}\bra{\Psi(N)}]$ we pick up a factor of
${}_{s,\ibar}\bra{\Phi'^{(N)}_{k'}} \Phi'^{(N)}_{k} \rangle_{s,\ibar} = \delta_{k',k}$ and
${}_{p}\bra{\np0-k'} \np0-k\rangle_{p} = \delta_{k',k}$ respectively, which yields the \textit{diagonal},
reduced density matrices
\be{rho:reduced}
\rho_p(N)         = \sum_{k=0}^{\np0}\,P_{k}^{(N)}\,\ket{\np0-k}_p\bra{\np0-k}, \qquad
\rho_{s,\ibar}(N) = \sum_{k=0}^{\np0}\,P_{k}^{(N)}\,\ket{\Phi'^{(N)}_{k}}_{s,\ibar}\bra{\Phi'^{(N)}_{k}},
\ee
with non-unit probabilities $P_{k}^{(N)}$ given by \Eq{Psi:Normalized}.
Thus, the underlying origin of the probabilities $P_{k}^{(N)}$ arises from the one-shot decoupling of the sequence of separably emitted Hawking signal/idler modes, which are however, each separably coupled to the
BH `pump' mode at each time step $N$. This is in agreement with physical approaches to BH particle production advocated in previous work \cite{Gerlach:1977,Mathur:2009,Bradler_Adami:2015}.

It is also interesting to note that the form of the probabilities $P_{k}^{(N)}$ given by \Eq{Psi:Normalized} are reminiscent of the initially seeded signal states considered by both
Alsing \cite{Alsing:2015} and Adami and Ver Steeg \cite{Adami_VerSteeg:2014}. That is, if instead of using the initial signal/idler vacuum state $\ket{0}_{s,\ibar}=\ket{0}_s\ket{0}_\ibar$, one considers the state with $n_{s0}$
initial particles in the signal mode $\ket{\psi(0)} = \ket{\np0}_p\,\ket{n_{s0}}_s\ket{0}_\ibar$
one would obtain in the short time limit $z<z^*$

\be{psi:soln:ns0}
|c^{<,(n_{s0})}_{k}(z,\tau)|^2 = p_{k}^{(\np0\rightarrow\infty)}=
  (1-z)^{n_{s0}+1} \, z^{k}
  \left(
         \begin{array}{c}
          k + n_{s0} \\
          k
        \end{array}
       \right),\,
\qquad 0 \le z \le z^*,
\ee
which is of the same form as $\tilde{P}_{k}^{(N)}$ in \Eq{Psi:Normalized} if we take $n_{s0} = N-1$.
The formal equivalence of the two probability distributions is the essential reason that the trilinear model considered by Alsing in \cite{Alsing:2015} with only one emitted Hawking signal/idler mode was capable to reproducing the Page information curves for long time evolution (see Fig.14 in \cite{Alsing:2015}), noting that the model should only be considered in the region where $dn_p(\tau)/d\tau\le0$ holds (and whose validity stops once $dn_p(\tau)/d\tau=0$ again with $\tau>0$).



\section{Numerical Results  and Finer Approximations to the Probabilities}\label{NumericalResults}
\subsection{Probabilities and entropy curves}
\begin{figure}[h]
\includegraphics[width=3.25in,height=2.0in]{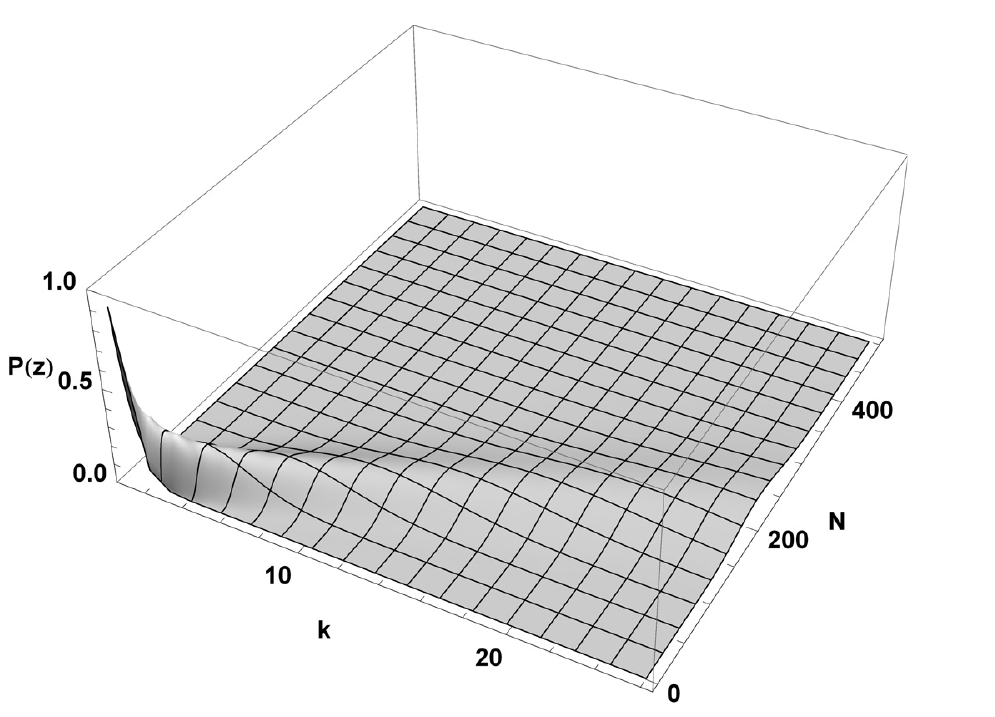}
\caption{Plot of $P^{(N)}_{k}(z)$, for $\np0=25$ and  $z = 0.1$.
}\label{fig:Probs}
\end{figure}

In Figure \Fig{fig:Probs} we plot
plot the probabilities $P^{(N)}_{k}(z)$ using $z=0.1$ for various values of $j_N\equiv k$, the collective number of particles emitted in all the signal/idler modes at iteration $N$.
%

\begin{figure}[h]
\includegraphics[width=3.0in,height=2.0in]{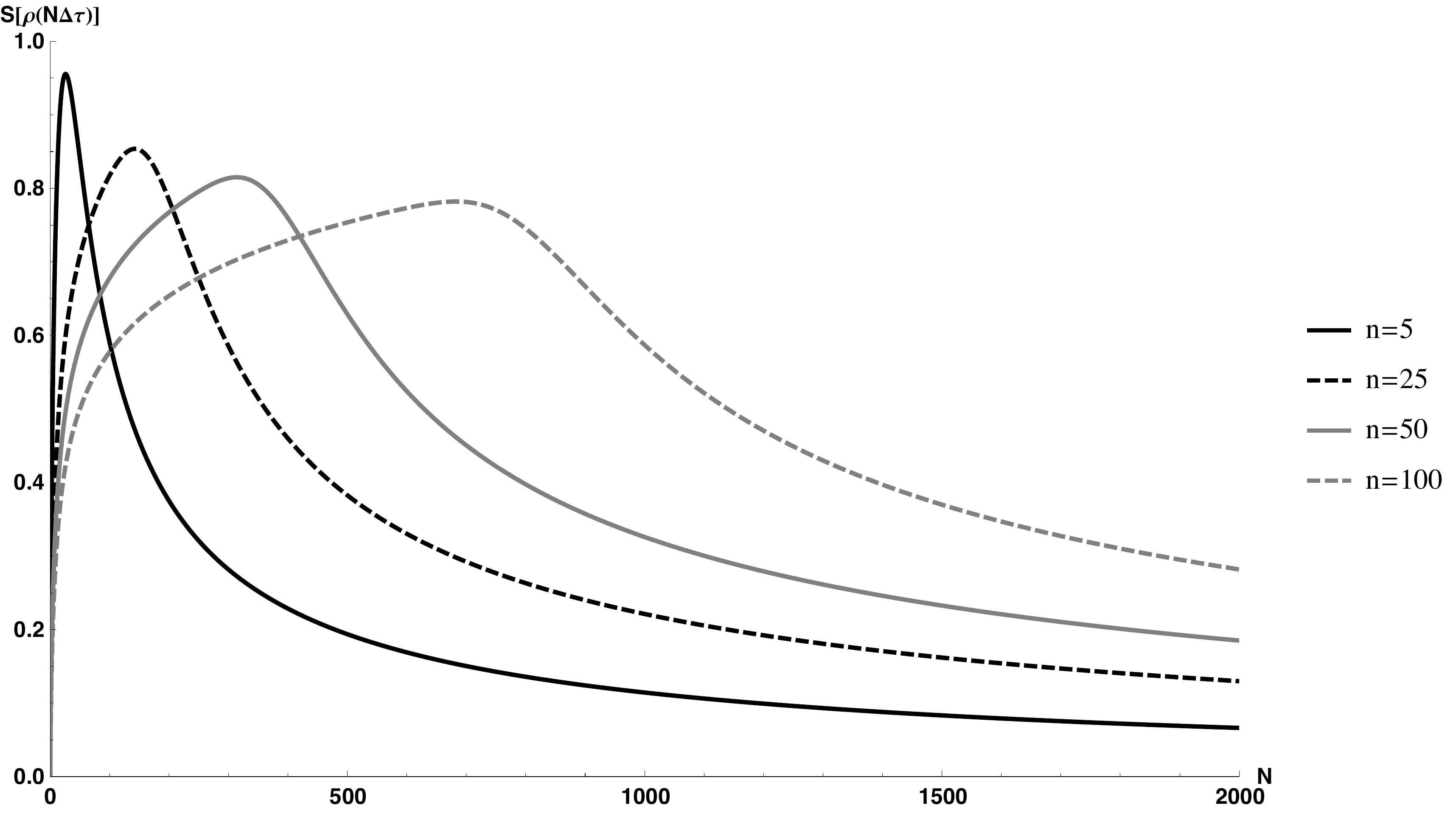}
\caption{$S_p^{(N)}(z)$ vs $N$ for  $\np0=(5,25,50,100)$ and $z=0.1$.
Entropies are computed with $\log_{\np0+1}$.
}\label{fig:Entropies}
\end{figure}

In \Fig{fig:Entropies}
 we plot the von Neumann entropy
$S_p^{(N)}(z)=S_{s,i}^{(N)}(z)
= -\sum_{k=0}^{\np0}\,P^{(N)}_{k}(z)\,\log_{\np0+1}\,P^{(N)}_{k}(z)$
for the reduced density matrix
$\rho_p^{(N)}(z)=Tr_{s,\ibar}[\rho_{p,s,\ibar}]$
or
$\rho_{s,\ibar}^{(N)}(z)=Tr_{p}[\rho_{p,s,\ibar}]$
for $\np0=(5,20,50,100)$ (as in \cite{Bradler_Adami:2015}) and $N=(1,\ldots,N_{max})$ with $N_{max} = (2000,4000,8000,10,000)$, after which we scale all graphs to $N_{max}=2000$.
The utilization of the logarithms  base $\np0+1$ is to ensure that the von Neumann entropies
remain less than unity, for comparison.
%
The entropy curves in \Fig{fig:Entropies} have the characteristic features of (i) an initial value of zero, appropriate for an initial product state at $N=0$, (ii) a rise to a peak value less than unity, and (iii) a tapering off towards zero for long-times (large $N$).

\begin{figure}[h]
\includegraphics[width=6.0in,height=3.0in]{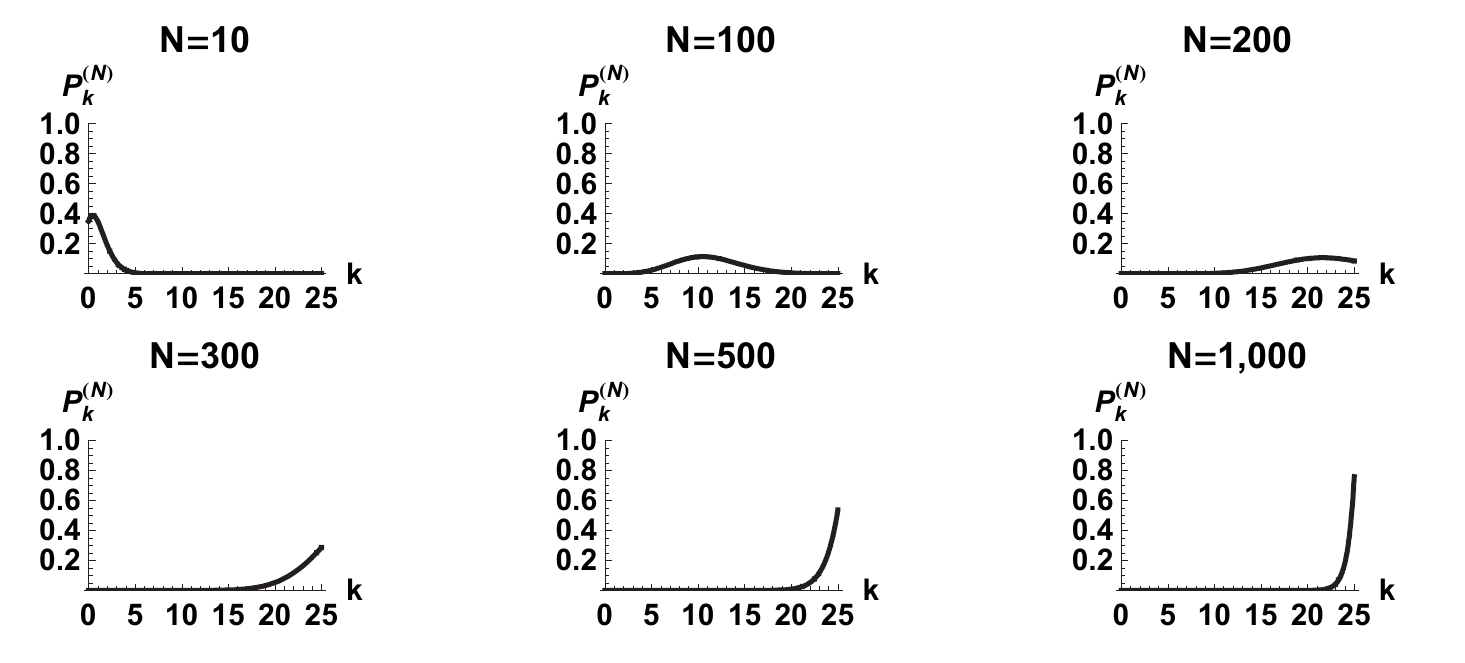}
\caption{$P_k^{(N)}(z)$ vs $k$ for $\np0=25$, $z=0.1$ and $N=(10,100,200,300,500,1000)$.
}\label{fig:Pksi:uncorrected:plots}
\end{figure}

An examination of the probabilities $P_k^{(N)}$ for various values of $N$ for $\np0=25$
\Fig{fig:Pksi:uncorrected:plots} reveals that they are peaked near $k=0$ for early times, when few Hawking particles are in the signal/idler modes and the state is essentially a separable product of squeezed states with the BH `pump' mode occupation number near its initial value of $\np0$. As time, i.e. $N$, evolves this peak moves steadily from low values of $k$ to high values of $k$, and for long times clusters about $k=\np0$, where the BH has essentially evaporated. In this longtime regime, we would expect BH `pump' to be in the state $\ket{\np0-k\approx0}_p$ with the emitted signal/idler field approximately in the state $\ket{\Phi'^{(N)}_{k\approx\np0}}$,
i.e. again, an approximate product state for $\ket{\Psi(N)}$.
As such, we would expect the entropy curves for longtimes to again be zero.
We address this in the next section by further approximating the probabilities $P_k^{(N)}$.

\subsection{Finer approximations to the probabilities}
We desire to further approximate the
the unnormalized state $\ket{\Phi^{(N)}_{j_N}}$ defined by the expression in the large square brackets in
\Eq{Psi:j:jN:first}. In particular, in the previous section we had approximated the terms
$(1-z^{\np0-j_{i-1}+1})^{-1/2}$ by unity. Recalling that $j_0=0$, we can factor out from all the nested sum an overall constant term $(1-z^{\np0+1})^{-1/2}\rightarrow 1$ for $z\ll 1$ and any reasonable sized value of $\np0$. The remaining factors have to be summed from $j_i=j_{i-1}:j_N\equiv k$, in succession from the inner summations, outwards. These complicated nested sums are what led to the numerical lattice-path approach of Br\'adler and Adami \cite{Bradler_Adami:2015}. Here, we make the simplified, but reasonable approximation that $(1-z^{\np0-j_{i-1}+1})^{-1/2}$, is dominated by its largest contribution $j_i=k$ from the upper limit of the summation, yielding $(1-z^{\np0-k+1})^{-1/2}$ which can then be factored out of all the nested summations, except the outermost one over $k$ itstelf. Since there are $N-1$ summations at time $N$ we obtain the slightly refined approximation to the probabilities
\be{Pksi:corrected}
\tilde{P}_{k}^{(N)} = (1-z)^N\, \frac{z^{k}}{(1-z^{\np0-k+1})^{N-1}}\,
\left(
       \begin{array}{c}
          k + N-1 \\
          k
        \end{array}
      \right),
\qquad
P_{k}^{'(N)} = \frac{\tilde{P}_{k}^{(N)}}{\sum_{k'=0}^{\np0}\,\tilde{P}_{k'}^{(N)} }.
\ee
\Eq{Pksi:corrected} constitutes the second primary analytical result of this present work.
\begin{figure}[h]
\begin{tabular}{cc}
\includegraphics[width=3.25in,height=2.0in]{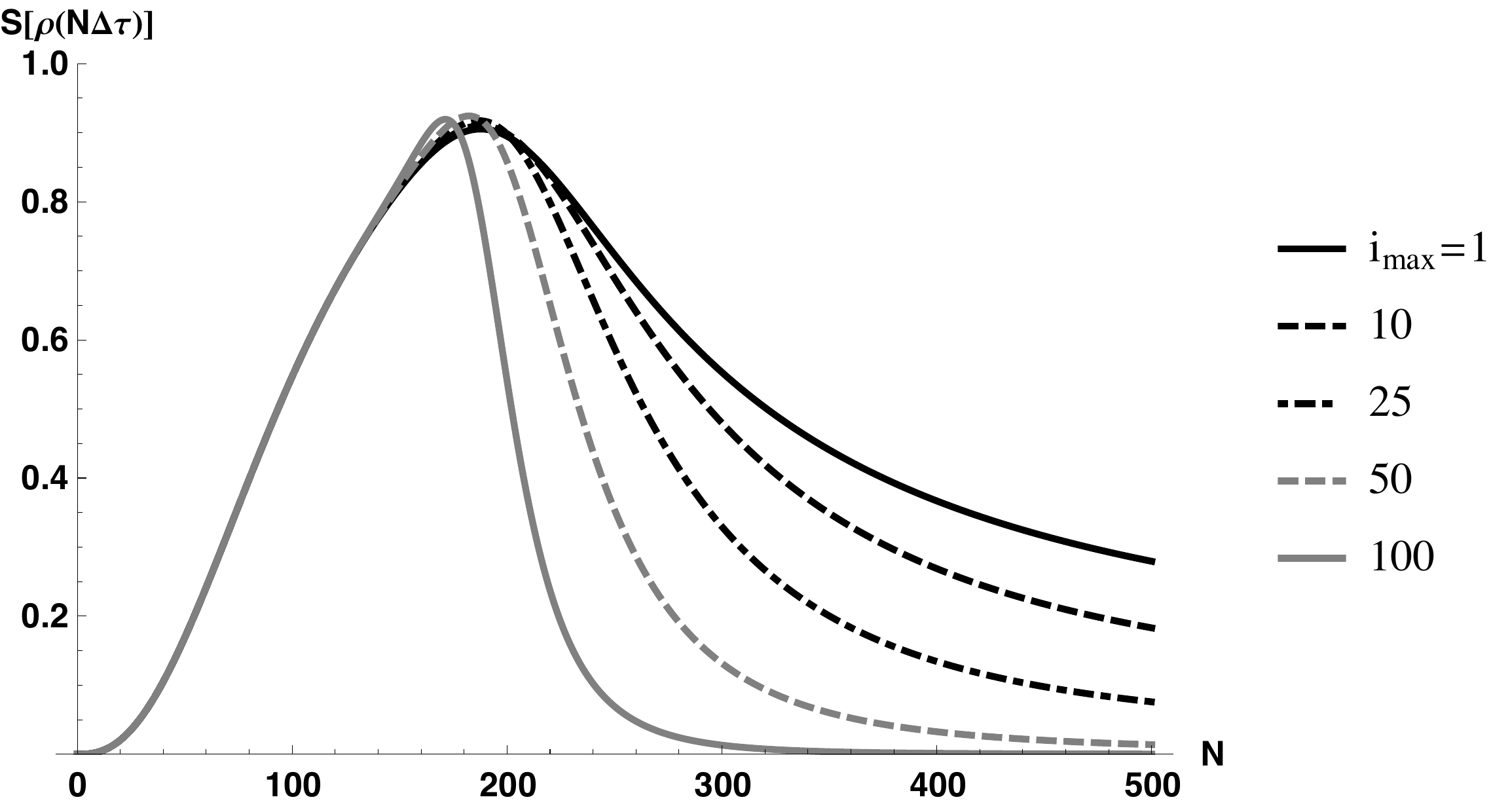}  &
\includegraphics[width=3.25in,height=2.0in]{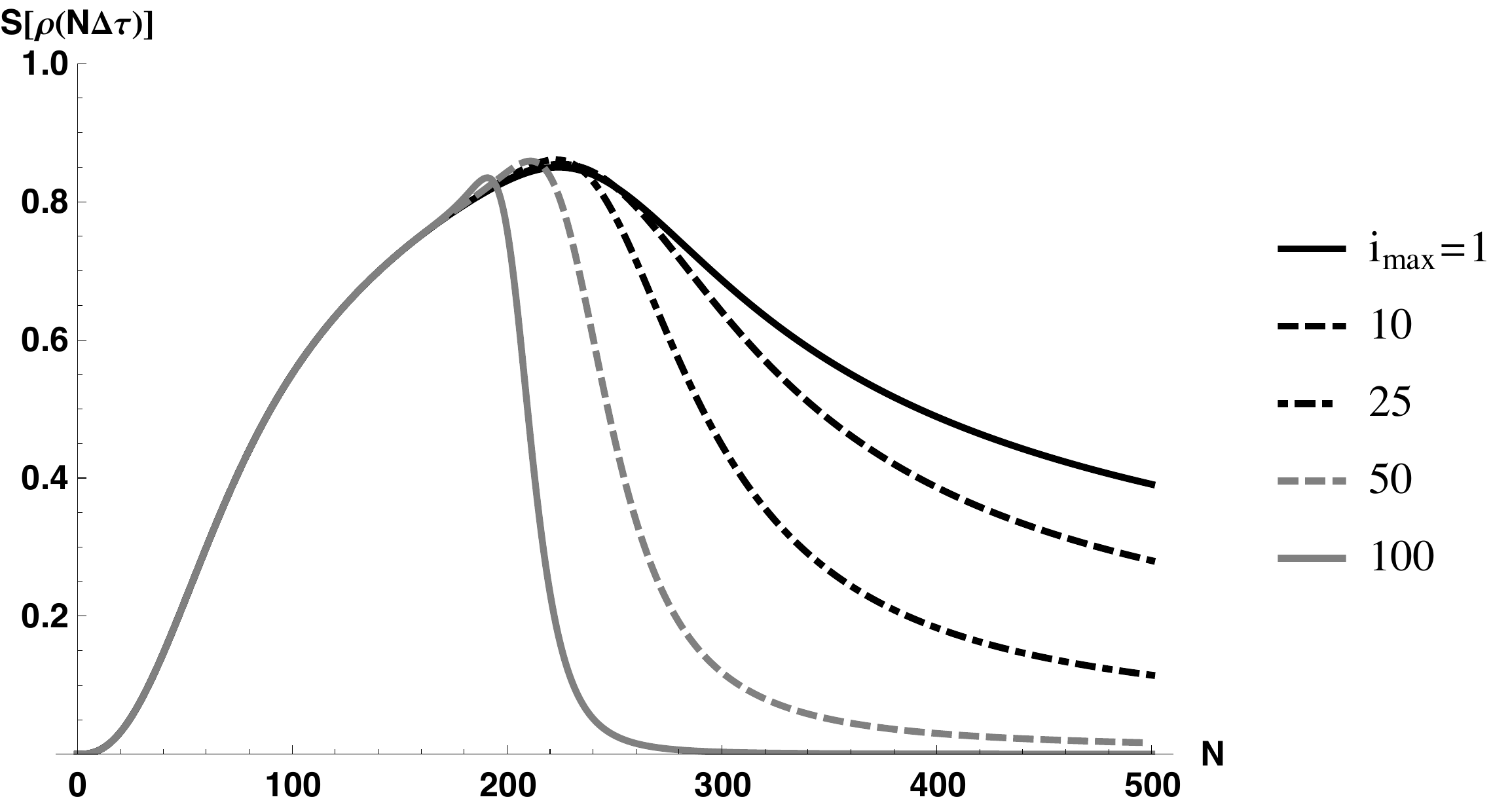}
\end{tabular}
\caption{Entropies with the extra term
$(1-z^{\np0-k+1})^{\textrm{min}(N-1,i_{max}-1)}$ in \Eq{Pksi:corrected} with various values of $i_{max}$ for
(left) $\np0=10$, (right) $\np0=25$.
Entropies are computed with $\log_{\np0+1}$.
}\label{fig:Pksi:imax}
\end{figure}
The effect of the extra factor $(1-z^{\np0-k+1})^{N-1}$ in $P_{k}^{(N)}$ can be seen by replacing it with
$(1-z^{\np0-k+1})^{\textrm{min}(N-1,i_{max}-1)}$ and varying the value of $i_{max} \le N$.
The value of $i_{max}$ sets how many terms $(1-z^{\np0-j_{i-1}+1})^{-1/2}$ in the $N-1$ nested sums in \Eq{Psi:j:jN:first} that we do not approximate as unity.
This is shown in \Fig{fig:Pksi:imax} for the cases of $\np0=10$ (left) and $\np0=25$ (right). These figures show how the additional factors
of  $(1-z^{\np0-k+1})$ in $P_{k}^{(N)}$ brings down the tail of entropy distribution $S$ to zero for longtimes, while leaving the short time (small $N$) portion of $S$ essentially unaltered.

\subsection{Page Information}
To investigate how the information emerges from the BH as it evaporates,
the \textit{Page information}, we follow Page's 1993 paper \cite{Page:1993b} (and \cite{Nation_Blencowe:2010,Alsing:2015}) and define the information $I$ as
\be{PageInfo}
I(\tau) = S_{thermal}(N) -  S\big(\rho_{s,\ibar}(N)\big).
\ee
Here $S_{thermal}$ is the effective thermal distribution $\rho_{thermal}(z_{thermal})$ with probability distribution
given by
$p_n^{thermal}=\left(1-z_{thermal}\right) z_{thermal}^n$ \Eq{psi:soln}, with
$z_{thermal}= \bar{n}_{s,\ibar}/(\bar{n}_{s,\ibar}+1)$, and
$\bar{n}_{s,\ibar}= \sum_{k=0}^{\np0}\, k\, P_k^{(N)}$
computed from $\rho_{s,\ibar}(\tau) = Tr_{p}\left[\ket{\Psi(N)}\bra{\Psi(N)}\right]$.
Lastly, from  \Eq{z:defn}, the (squeezing) rapidity $z$ is defined as $z = \tanh^2(\tau)$ with $\tau = \sqrt{\np0}\,r\,t$.
Taking $t_N = N \Delta t = N/N_{max}$ for $N=1:N_{max}$  we utilize (with $r\equiv 1$)
$z =  \tanh^2\left[\tanh^{-1}(z_{max})\,\tanh(\sqrt{\np0}\,N/N_{z_{max}})\right]$
with $0<z_{max}<1$ where $N_{z_{max}}\le N_{max}$
determines a controllable build-up time from $z=0$ to $z=z_{max}$.
Since each unitary $\{U_{p,i}\}|_{i=1:N}$
acts over a small time, we allow the rapidity to gradually build up from $z=0:z_{max}\ll 1$, assuming weak squeezing in the emitted Hawking signal/idler radiation pairs by the BH.
(Note: similar curves are obtained by simply setting $z=z_{max}$ for all $N$).
\begin{figure}[h]
\begin{tabular}{cc}
\includegraphics[width=3.65in,height=2.0in]{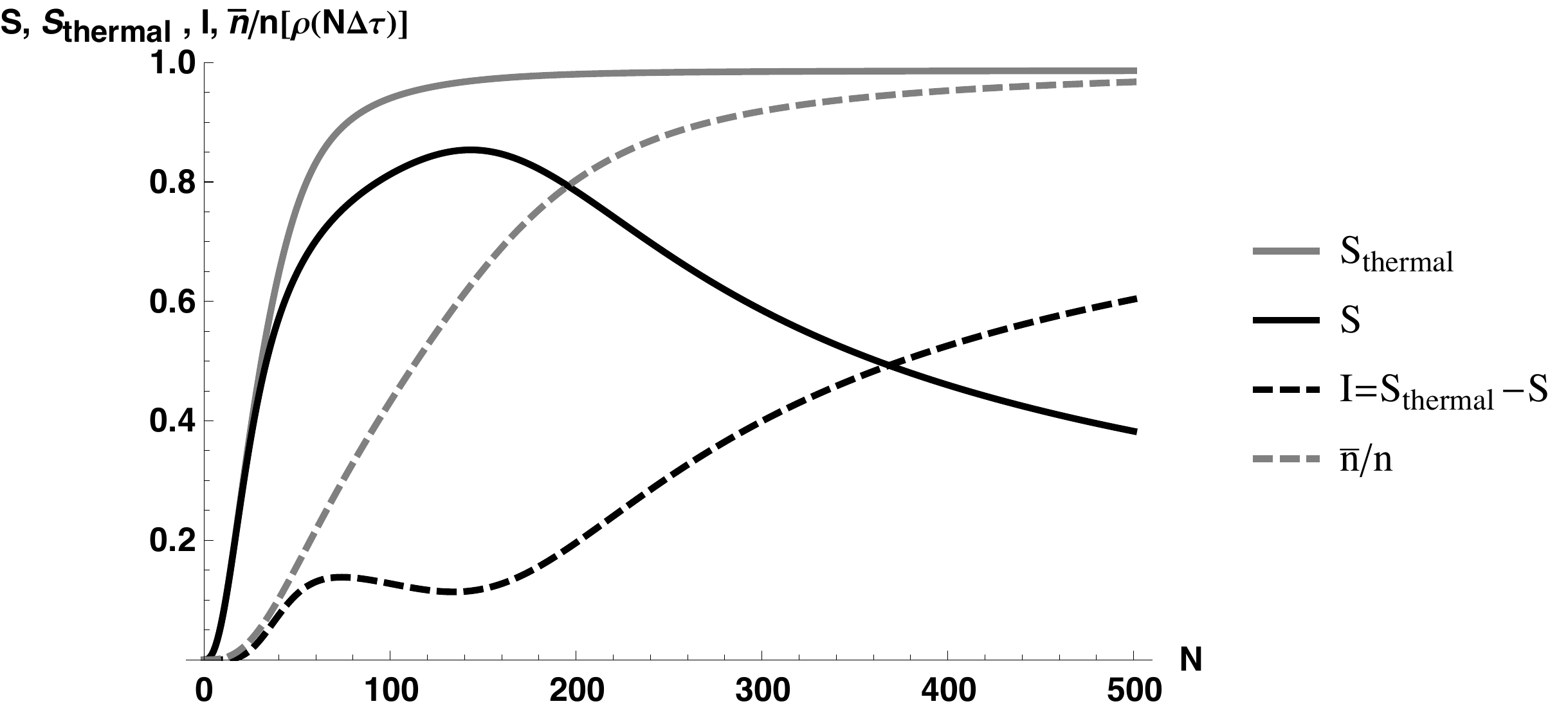}  &
\includegraphics[width=3.65in,height=2.0in]{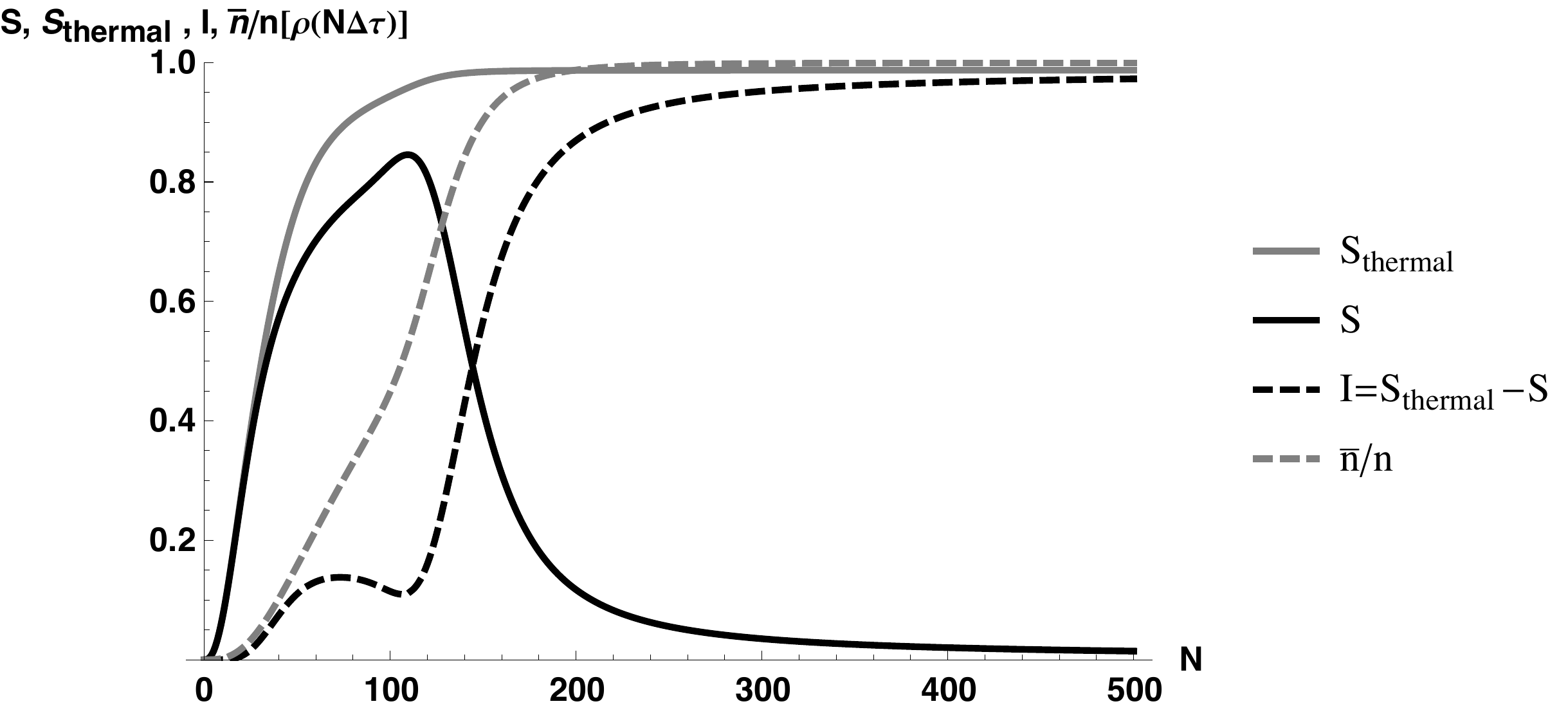}
\end{tabular}
\caption{Plots of entropy $S$ (black, solid), effective $S_{thermal}$ (gray, solid), Page Information $I$ (black, dashed) and total number of emitted Hawking particles in all signal/idler modes $\bar{n}_{s,\ibar}/\np0$ vs time $N$
for $\np0=25$ and $z_{max}=0.1$, ($N_{z_{max}}=200, i_{max}=50$),
using probabilities (left) $P_k^{(N)}$ in \Eq{Psi:Normalized}, and (right) $P_k^{'(N)}$ in \Eq{Pksi:corrected}.
Entropies are computed with $\log_{\np0+1}$.
}\label{fig:Allcurves:n25}
\end{figure}

In \Fig{fig:Allcurves:n25} we show plots of the entropy $S$ (black, solid), effective $S_{thermal}$ (gray, solid), Page Information $I=S_{thermal}-S$ (black, dashed) and total number of emitted Hawking particles in signal/idler modes $\bar{n}_{s,\ibar}/\np0$ vs time $N$
for $\np0=25$
using probabilities (left) $P_k^{(N)}$ in \Eq{Psi:Normalized}, and (right) $P_k^{'(N)}$ in \Eq{Pksi:corrected}.
Entropies are computed with $\log_{\np0+1}$ so that all graphs have maximum value of unity, for comparison.
Both curves show that for early times (small $N$) $S\approx S_{thermal}$ so that the Page information $I$ is flat with, with very small slope. As time progresses, $I$ begins to grow, as $\bar{n}_{s,\ibar}$ rapidly increases, and the BH begins to evaporate. For $\bar{n}_{s,\ibar}/\np0 > 1/2$ there are less particles in the BH `pump' mode than have been emitted into all the Hawking radiation signal/idler modes and $S$ begins to decrease.
Using $P_k^{(N)}$ from \Eq{Psi:Normalized} $S$ in  \Fig{fig:Allcurves:n25}-(left) decreases very slowly as $\ket{\Psi(N)}$ approaches a nearly separable state $\ket{\np0-k\approx 0}_p\,\ket{\Phi^{(N)}_{k\approx\np0}}_{s.\ibar}$, and the information slowly saturates to a value of unity.
In \Fig{fig:Allcurves:n25}-(right) the effect of the refined probabilities $P_k^{'(N)}$ from \Eq{Pksi:corrected}
reduce $S$ more rapidly to zero, appropriate for a final separable state, with a commensurate faster saturation of the information to unity.

\begin{figure}[h]
\includegraphics[width=5.0in,height=2.0in]{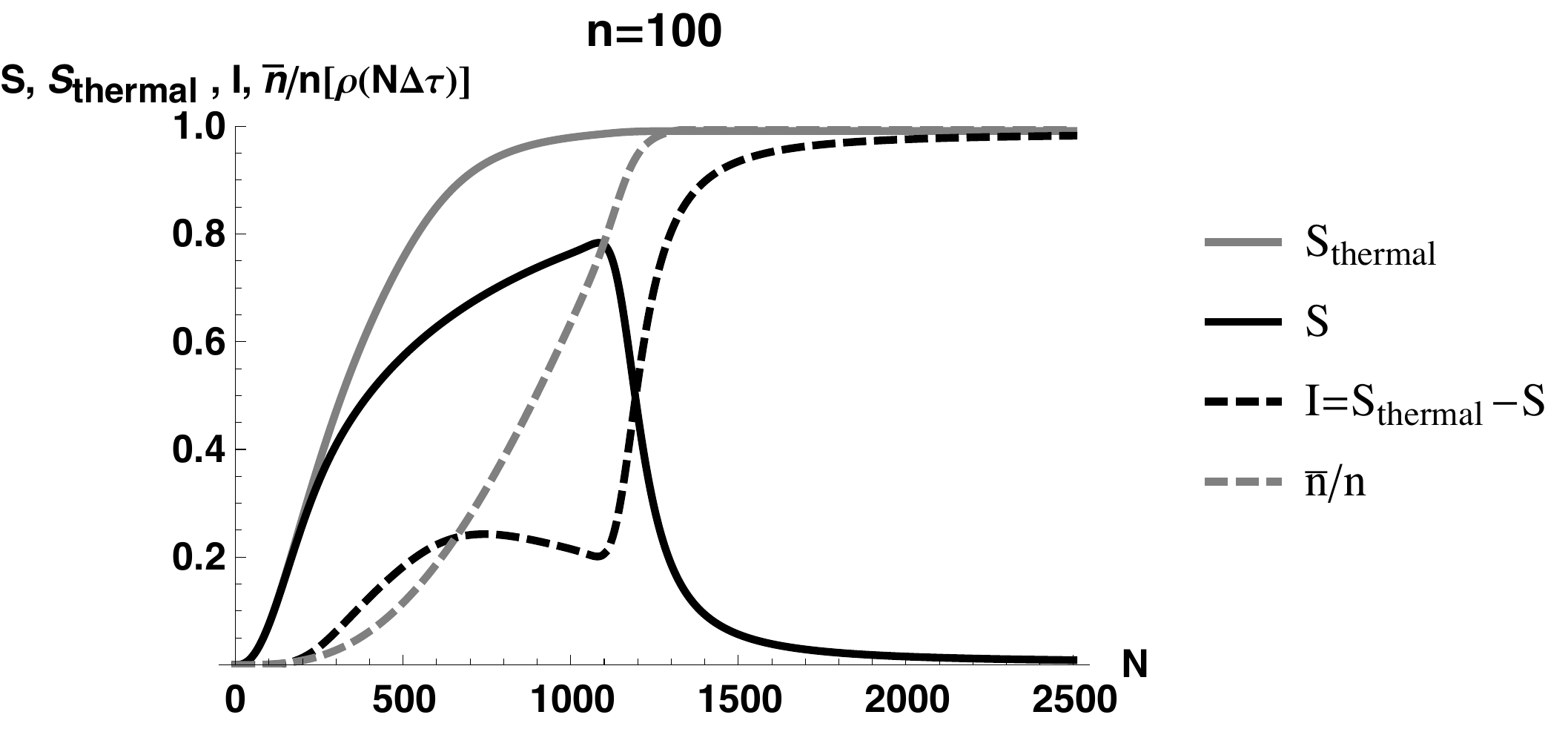}
\caption{Plots of entropy $S$ (black, solid), effective $S_{thermal}$ (gray, solid), Page Information $I$ (black, dashed) and total number of emitted Hawking particles in signal/idler modes $\bar{n}_{s,\ibar}/\np0$ vs time $N$
for $\np0=100$ and $z_{max}=0.1$, ($N_{z_{max}}=10^4, i_{max}=50$),
using probabilities $P_k^{'(N)}$ in \Eq{Pksi:corrected}.
Entropies are computed with $\log_{\np0+1}$.
}\label{fig:Allcurves:n100}
\end{figure}

In \Fig{fig:Allcurves:n100} we show the similar entropy and information curves for $\np0=100$, where the initial flatness of the information $I$ is more pronounced.
\Fig{fig:Allcurves:n25}-(right) and \Fig{fig:Allcurves:n100}  constitutes the main results of this present work.
Note that a brute force summation of all the terms in \Eq{Psi:j:jN:first} would involve the addition of
on the order of
$
\small{
\left(
       \begin{array}{c}
          k + N-1 \\
          k
        \end{array}
      \right)
}
$
summands, which equates to $10^{42}$ and $10^{182}$ terms for $k=\np0=25,\,N=500$ and $k=\np0=100,\,N=2500$ for
\Fig{fig:Allcurves:n25}-(right) and \Fig{fig:Allcurves:n100} respectively, which is impractical.
While most of the summands would be negligibly small to warrant approximating to zero,
a reasonable estimate of only $k=10$ nonzero terms per sum would still
lead to the prohibitive total number of nonzero summands of
$10^{20}$ and $10^{27}$ for $N=500$ and $N=2500$, respectively.
Hence, the necessity for the analytic approximations to the probabilities given
by \Eq{Psi:Normalized} and \Eq{Pksi:corrected}.

\begin{figure}[h]
\begin{tabular}{cc}
\includegraphics[width=2.5in,height=1.75in]{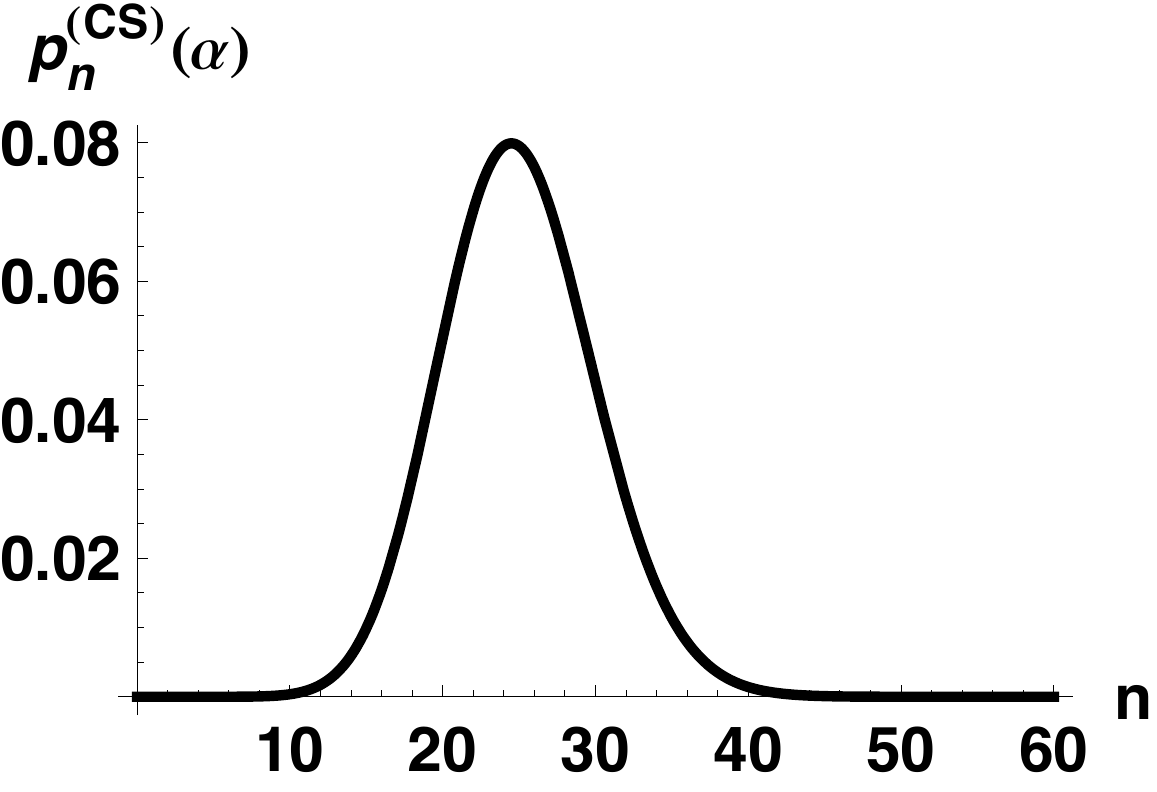}  &
\includegraphics[width=3.75in,height=1.75in]{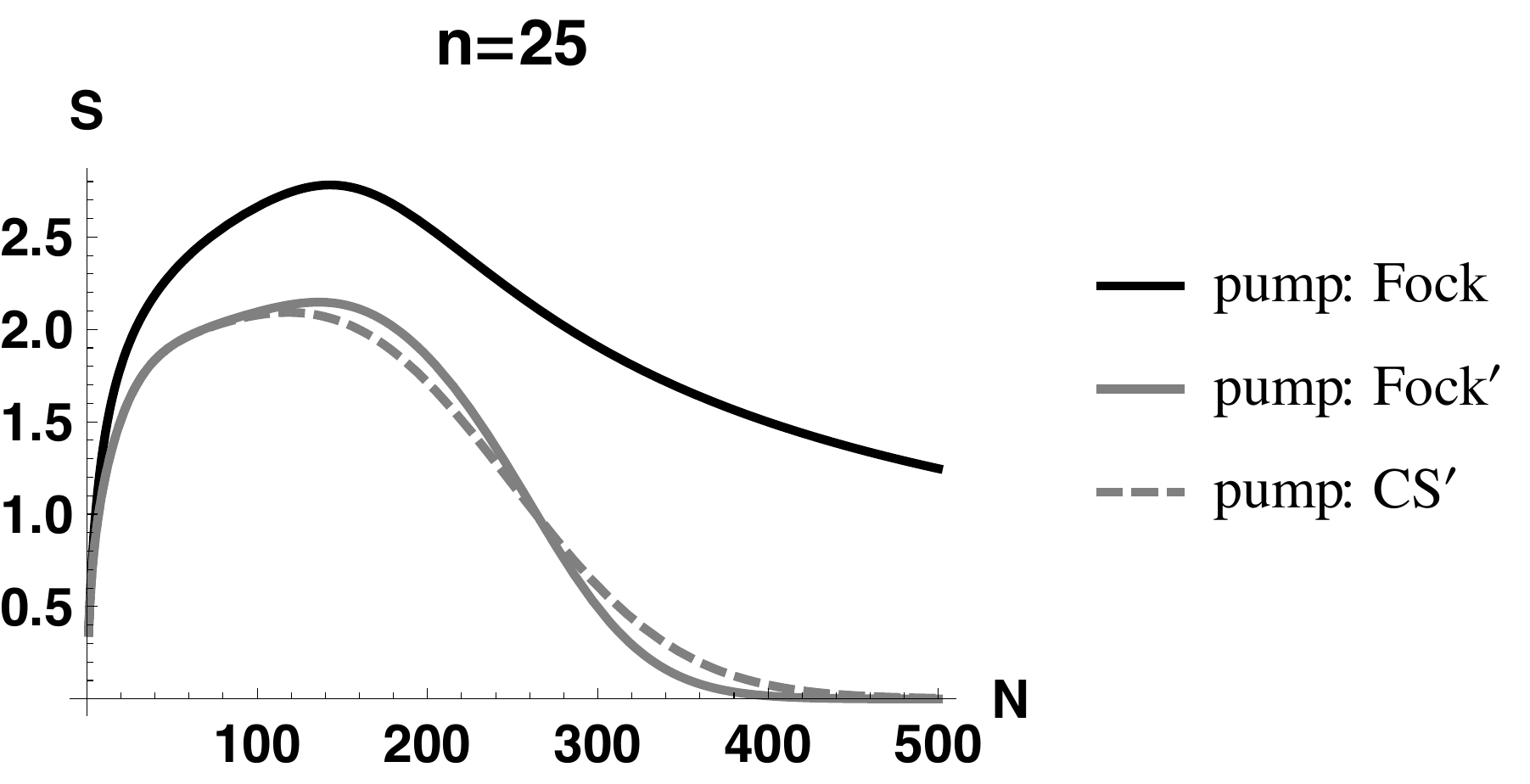}
\end{tabular}
\caption{(left) Initial coherent state probabilities $p_n^{(CS)}(\alpha)$ for $\alpha^2=25$,
(right) entropy curves $S$ for probabilities
(black, solid) $P_k^{(N)}$ in \Eq{Psi:Normalized},
(gray, solid) $P_k^{'(N)}$ in \Eq{Pksi:corrected}, and
(gray, dashed) $p_n^{(CS)}(\alpha)\, P_k^{'(N)}$.
Entropies are computed with $\ln$.
}\label{fig:S:CS:n25}
\end{figure}

As explored in Alsing \cite{Alsing:2015}, the intial BH `pump' might also be represented by a coherent state
$\ket{\alpha} = e^{-|\alpha|^2/2}\,\sum_{n=0}^\infty \alpha^n \ket{n}/\sqrt{n!}$ such that $a_p\ket{\alpha} = \alpha\ket{\alpha}$, the quantum state that most approximates a classical state with mean particle number $\alpha^2=\np0$.
This leads to the state
$
\ket{\Psi(N)}~\approx~
\sum_{\np0=0}^\infty\, \sqrt{p_{\np0}^{(CS)}(\alpha)}\,\sum_{k=0}^{\np0} \sqrt{P_{k}^{'(N)}}\, \,\ket{\np0-k}_p \,\ket{\Phi'^{(N)}_{k}},
$ with probabilities $P_{k}^{'(N,CS)} = p_{\np0}^{(CS)}\,P_{k}^{'(N,\np0)}$
\footnote{The reduced density matrix for the signal/idler (and BH `pump') field is in general non-diagonal and given by
$\rho_{s,\ibar} =
\sum_{\np0=0}^\infty\,\sum_{\np0'=0}^\infty\,
\sqrt{p_{\np0}^{(CS)}(\alpha)\,p_{\np0'}^{(CS)}(\alpha)}\,
\sum_{k=0}^{\np0} \sqrt{P_{k}^{'(N,\np0)} P_{k}^{'(N,\np0)}}\, \ket{\Phi'^{(N,\np0)}_{k}}  \bra{\Phi'^{(N,\np0')}_{k}}
$
which we can approximate by its diagonally dominant $\np0'=\np0$ contributions as
$
\rho_p\approx \sum_{\np0=0}^\infty\,$
$\sum_{k=0}^{\np0}\,p_{\np0}^{(CS)}(\alpha)\,P_{k}^{'(N,\np0)},
\ket{\Phi'^{(N,\np0)}_{k}}  \bra{\Phi'^{(N,\np0)}_{k}}
$
to illustrate the desired smoothing effect.
} 
which tends to smooth out the entropy curve $S$ as shown in \Fig{fig:S:CS:n25}-(right) by averaging original probabilities $\,P_{k}^{'(N)}$ over the initial coherent state probability distribution $ p_{\np0}^{(CS)}$. The main point is that $P_{k}^{'(N,CS)}$ is well approximated by $\,P_{k}^{'(N)}$ with $\alpha^2=\np0$.
%
\section{Discussion}\label{Discussion}
\begin{figure}[h]
\includegraphics[width=4.0in,height=3.0in]{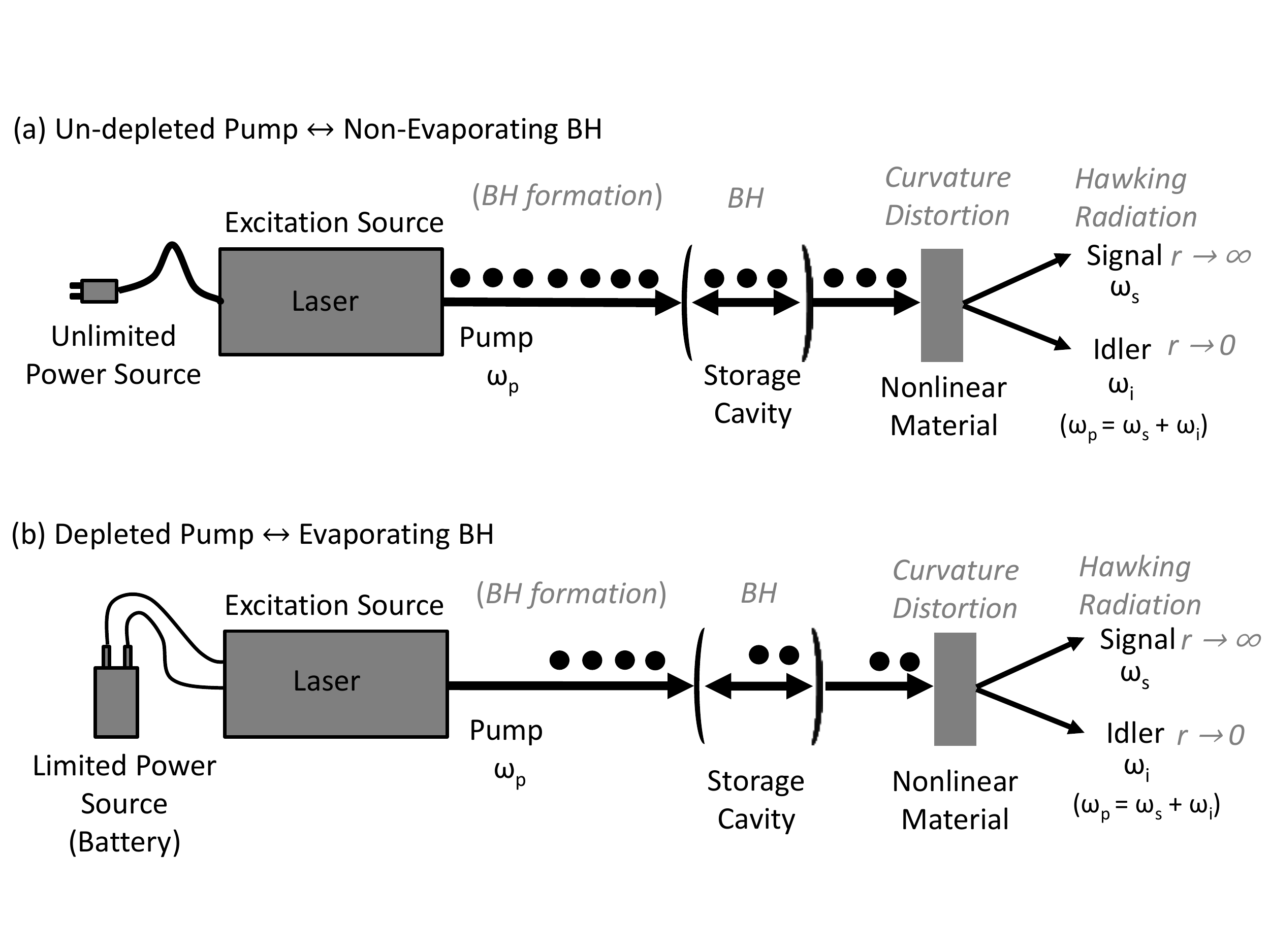}
\caption{Analogy of BH evaporation to SPDC process.
}\label{fig:BH:SPDC:analogue:fig1}
\end{figure}

Here we propose that the process of BH evaporation has a very strong analogy to the process of spontaneous parametric down conversion (SPDC) \cite{Gerry:2004,Agarwal:2013} with the emitted Hawking radiation acting as the spontaneously  signal/idler pairs generated by the trilinear Hamiltonian \Eq{H:eqn} driven by the BH modeled as a depleted pump laser source.
In \Fig{fig:BH:SPDC:analogue:fig1}(a) we illustrate a laser excitation source powered by an unlimited power source (e.g. A/C wall socket) driving an optical storage cavity. This optical cavity, with non-zero mirror transmission subsequently
pumps a non-linear crystal in which a pump photon of frequency $\omega_p$ spontaneously down-converts two photons from the vacuum, the signal at frequency $\omega_s$ and the idler at frequency $\omega_\ibar$ such that energy is conserved,
$\omega_p = \omega_s + \omega_\ibar$. As long as the power source is unlimited, signal/idler pairs are continually created %
%
as squeezed state signal/idler pairs, and the pump can be treated as quantized, though with large enough occupation number $\np0$ to essentially be treated as a (non-depleting) constant. This is the analogy of a non-evaporating BH generating purely thermal Hawking radiation. The laser excitation source filling the storage cavity can loosely be thought of as the analogy of the BH formation, while the storage cavity can be thought of as some region of finite width about the BH horizon (see Discussion in \cite{Alsing:2015}).

\Fig{fig:BH:SPDC:analogue:fig1}(b) shows essentially the same set up as in \Fig{fig:BH:SPDC:analogue:fig1}(a) except for one crucial difference: the initial laser excitation source is driven by a power source of limited energy, here illustrated as a battery. Again the BH formation fills up the storage cavity, but now with a large but \textit{finite} number $\np0$ of quantized particles which can still drive the SPDC Hawking radiation production process. In this later case, the storage cavity will eventually deplete itself (e.g. finite battery life). The signal/idler pairs will still be produced as essentially squeezed states at each SPDC excitation, with each emitted pair entangled with the BH `pump' source at the given time step $N$ (but not at subsequent time steps). The analogy of the non-linear crystal responsible for the SPDC generation of Hawking signal/idler pairs is the curvature distortion near the BH horizon, as discussed in \cite{Boulware:1976,Gerlach:1977,Mathur:2009}.

\begin{figure}[h]
\includegraphics[width=3.5in,height=2.5in]{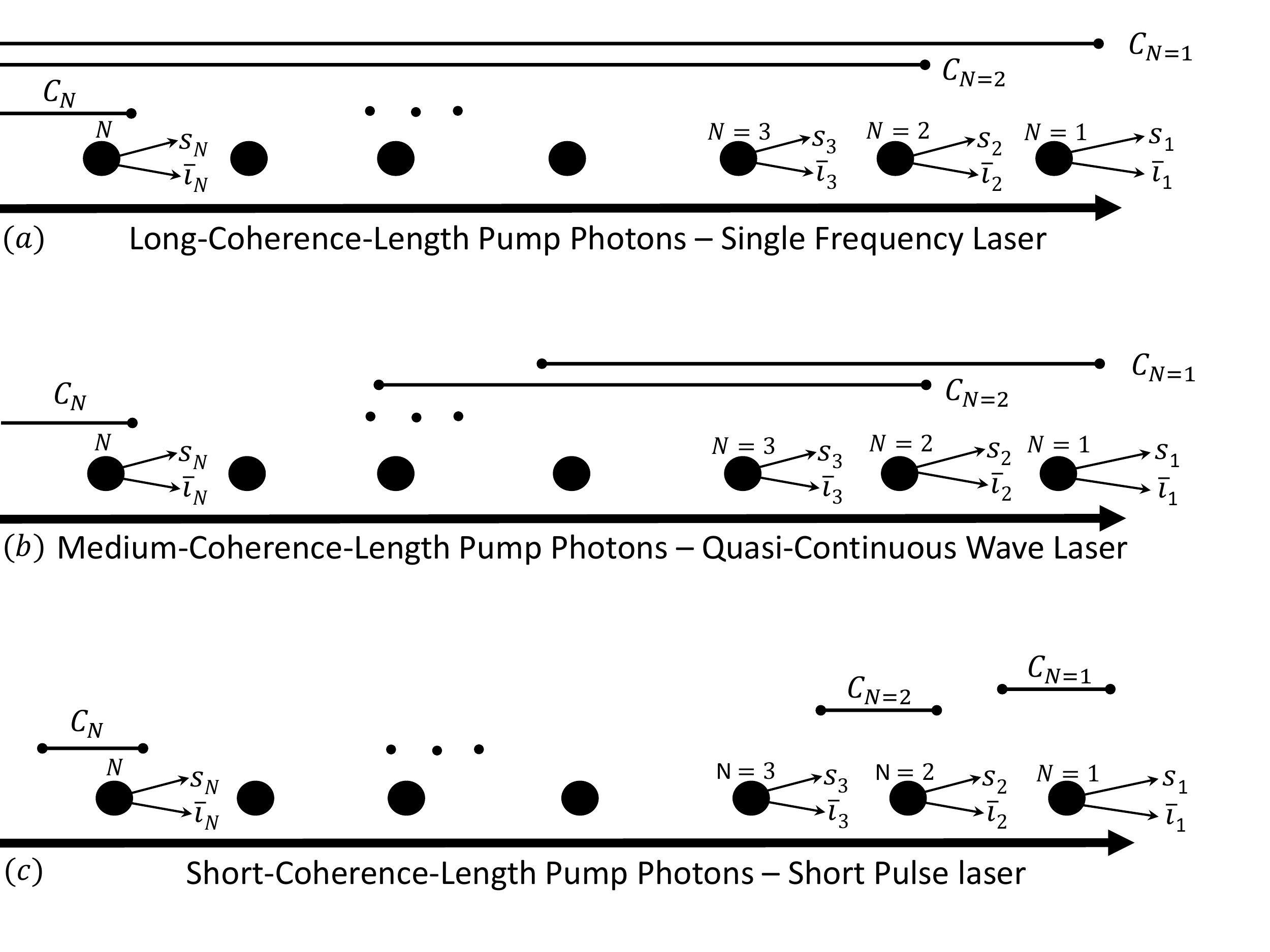}
\caption{Consideration of coherence length of BH `pump' source particles.
}\label{fig:BH:SPDC:analogue:fig2}
\end{figure}
By simply writing down a pure state wavefunction $\ket{\Psi(N)}$ in \Eq{Psi:N} to describe our quantum state, one has implicitly assumed some level of coherency of the pump source which in turns effects the level of entanglement of the emitted signal/idler particles with the BH `pump' particles, which without loss of generality, and for comparison with our laser SPDC analogy, we treat in this discussion as photons. This is illustrated in \Fig{fig:BH:SPDC:analogue:fig2}.
In \Fig{fig:BH:SPDC:analogue:fig2}(a) we illustrate a pump laser source of infinite
coherence length $C_N \sim c/\Delta\omega_p$. Here, $\Delta\omega_p$ is the bandwidth of BH `pump' photons
which for \Fig{fig:BH:SPDC:analogue:fig2}(a) is considered very narrow, $\Delta\omega_p \ll \omega_p$. Consequently, all signal/idler SPDC pairs generated are coherent with the pump simultaneously, and therefore, indirectly with each other, causing various degrees of entanglement as explored by Alsing \cite{Alsing:2015}. This is the case considered by Nation and Blencowe \cite{Nation_Blencowe:2010} and Alsing \cite{Alsing:2015}, which can also be equivalently considered as having the non-linear material/curvature distortion occurring within the storage cavity.
As the coherence length of the BH pump photons decreases ($\Delta\omega_p$ increases) \Fig{fig:BH:SPDC:analogue:fig2}(b), less and less SPDC generated pairs are coherent with with the pump, and hence indirectly with each other. If the coherence length of the pump photons is less than the average time $\Delta\tau$ ($\Delta N =1$ in our model) for SPDC signal/idler pair generation, as illustrated in \Fig{fig:BH:SPDC:analogue:fig2}(c), each generated SPDC signal/idler pair is only coherent with the BH `pump' until the next pair is generated. This is the one-shot decoupling model of Bradler and Adami \cite{Bradler_Adami:2015}, and explored analytically in this present work. As discussed earlier, this model is commensurate with physical models for Hawking radiation production \cite{Boulware:1976,Gerlach:1977,Mathur:2009} in which the curvature distortion at the BH horizon (of shrinking radius) creates the signal/idler pairs, which then propagate away from the region of generation, since the BH horizon decreases, shifting the region of curvature distortion for subsequent pair production.

As evidenced by the finer approximation utilized in  the previous section to produce the probabilities $P_k^{'(N)}$ in \Eq{Pksi:corrected}, the terms $\left.(1-z^{\np0-j_i+1})^{-1/2}\right|_{i=1:N}$ in the temporally nested sum in
$\ket{\Phi^{(N)}_{j_N}}$ \Eq{Psi:N} describe the detailed probability structure
of the total BH/Hawking radiation state $\ket{\Psi(N)}$.
Hence, further refinements than simply considering their maximum contribution from $j_i=k$ are warranted, yet difficult to implement analytically due to the large number of temporally ordered nested sums. Thus, it appears that some blend of an analytical approach detailed here, coupled with a numerical approach, as advocated by the lattice path method Br\'adler and Adami \cite{Bradler_Adami:2015} might prove fruitful in gaining further physical insight the nature of the one-shot decoupling bipartite state $\ket{\Psi(N)}$, and subsequently the detailed nature of the Page information of the Hawking radiation from an evaporating BH.

Considering extensions of this present work, we note that our model utilized a monochromatic, single frequency  BH `pump' source to drive the signal/idler generation i.e. a delta function frequency distribution for the pump.  A more general model, which might encapsulate a more realistic physical scenario, would be to model the BH `pump' source not as a single frequency, but as a collection of frequencies over some bandwidth, ie a pulse source.  Our model also described the emitted Hawking radiation signal/idler pairs as single frequency modes, which is analogous to a long non-linear crystal with strict phase matching conditions \cite{Gerry:2004,Agarwal:2013}.  A more general model could incorporate a frequency bandwidth for the emitted Hawking radiation signal/idler pairs, similar to a shorter non-linear crystal with a less restrictive phase matching condition.

\acknowledgments PMA would like to acknowledge  P.R. Rice and N. Natek for useful discussions.
The authors would like to acknowledge support for this work by T. Curcic of the Air Force Office
of Scientific Research (AFOSR).
Any opinions, findings and conclusions or recommendations expressed in this material are those of the author(s) and do not necessarily reflect the views of AFRL.



\begin{thebibliography}{99}
%
\bibitem{Nation_Blencowe:2010} P.D. Nation and M.P. Blencowe, \textit{The Trilinear Hamiltonian: a zero dimensional model of Hawking Radiation from a quantized source}, New J. Phys. \textbf{12}, 095013 (2010).
%
\bibitem{Alsing:2015} P.M. Alsing, "Parametric down conversion with a depleted pump as a model for classical information transmission capacity of quantum black holes," Class. Quantum Grav. \textbf{32}, 075010 (2015).
%
\bibitem{Bradler_Adami:2015} K. Br\'{a}dler and C. Adami, "One-shot decoupling and Page curves from a dynamical model for black hole evaporation," arXiv:1505.0284.
%
\bibitem{Page:1993a} D.N. Page, \textit{Average entropy of a subsystem}, Phys.Rev.Lett. \textbf{71},  1291-1294 (1993); arxiv:gr-qc:9305007v2.
%
\bibitem{Page:1993b} D.N. Page, \textit{Information in black hole radiation}, Phys.Rev.Lett. \textbf{71}, 3743-3746, (1993); arxiv:gr-qc:9306083v2.
%
\bibitem{Boulware:1976} D.G. Boulware, \textit{Hawking radiation and thin shells}, Phys. Rev. D.
\textbf{13}, 2169 (1976).
%
\bibitem{Gerlach:1977} U.H. Gerlach \textit{The mechanism of blackbody radiation from an incipient black hole},
Phys. Rev. D, \textbf{14}, 1479 (1976).
%
\bibitem{Mathur:2009} S.D. Mathur, \textit{The information paradox: a pedagotical introduction}, Class. Quantum Grav. \textbf{26}, 224001.
%
\bibitem{Gerry:2004} C. Gerry and P.L. Knight, \textit{Introductory Quantum Optics}, Cambridge Univeristy Press,
Cambridge (2004).
%
\bibitem{Agarwal:2013} G.S. Agarwal, {\textit{Quantum Optics}}, Cambridge Univ. Press (2013).
%
\bibitem{Walls:1970} D.F. Walls and R. Barakat, \textit{Quantum-Mechanical Amplification and Frequency Conversion with a Trilinear Hamiltonian}, Phys. Rev. A \textbf{1}, 446 (1970).
%
\bibitem{Bonifacio:1970} R. Bonifacio and G. Preparata \textit{Coherent spontaneous emission}, Phys. Rev. A \textbf{2}, 336 (1970).
%
\bibitem{Hawking:1975} S.W. Hawking \textit{Particle creation by black holes}, Comm. Math. Phys. \textbf{43} 199 (1975).
%
\bibitem{Unruh:1976} W.G. Unruh \textit{Notes on black hole evaporation}, Phys. Rev. D \textbf{14}, 870 (1976).
%
\bibitem{Yurke_Potasek:1987}B. Yurke and M. Potasek, \textit{Obtainment of thermal noise from a pure quantum state},
Phys. Rev. A \textbf{36}, 3464 (1987).
%
\bibitem{Stojkovic:2006:2015} A. Saini and D. Stojkovic, \textit{Radiation from a collapsing object is manifestly covariant}, arxiv:1503.01487v3; T. Vachaspati, D. Stojkovic and L.M. Krauss, \textit{Observation of incipient black holes and the information loss problem}, arxiv:gr-qc/0609024v3
%
\bibitem{Alberghi:2001} G.L. Alberghi, R. Casadio, G.P. Vacca and G. Venturi, \textit{Gravitational collapse of a radiating shell}, arxiv:gr-qc/0102014v2.
%
\bibitem{Pedrosa:1987:1989} I.A. Pedrosa, \textit{Canonical transformations and exact invariants for dissipative systems}, J. Math. Phys. A \textbf{28} 2662 (1987); \textit{ibid}, \textit{Coherent states for certain time-dependent systems}, Revista Brasileira de Fisicia \textbf{19}, 502 (1989).
%
\bibitem{Alsing_Rice:2015} P.M. Alsing and P.R. Rice, (\textit{in preparation}).
%
\bibitem{Adami_VerSteeg:2014} C. Adami and G. Ver Steeg, Class. Quantum Grav. \textbf{31}, 075015 (2014).
%
\end{thebibliography}
\end{document}